\documentclass[fleqn,usenatbib]{mnras}
\usepackage[utf8]{inputenc}
\usepackage{hyperref}
\usepackage{amsfonts}
\usepackage{amsmath,amssymb}
\usepackage{graphicx}
\usepackage{booktabs} 
\usepackage{appendix}
\usepackage{tabularx}
\usepackage{blindtext}
\usepackage{multicol}
\usepackage{mathtools}
\usepackage{xspace}
\usepackage{fix-cm}
\usepackage{float}
\DeclareRobustCommand{\VAN}[3]{#2}
\let\VANthebibliography\thebibliography
\def\thebibliography{\DeclareRobustCommand{\VAN}[3]{##3}\VANthebibliography}

\newcommand\rev[1]{{\color{black}#1}}
\newcommand\revnew[1]{{\color{black}#1}}

\newcommand\reltrans{\textsc{reltrans}\xspace}

\newcommand\xspec{\textsc{xspec}\xspace}
\newcommand\reltransdcp{\texttt{reltransDCp}\xspace}
\newcommand\rtdist{\texttt{rtdist}\xspace}

\newcommand\tbabs{\texttt{TBabs}\xspace}
\newcommand\diskbb{\texttt{diskbb}\xspace}
\newcommand\bbody{\texttt{bbody}\xspace}
\newcommand\xray{X-ray\xspace}

\title[Distance Measurement of Cygnus X-1]{X-ray reverberation black hole mass and distance estimates of Cygnus X-1}
\author[P. O'Neill]{Patrick O'Neill$^{1}$\thanks{E-mail: p.o'neill3@ncl.ac.uk}, Adam Ingram$^{1}$\thanks{E-mail: adam.ingram@newcastle.ac.uk}, Edward Nathan$^{2}$, Guglielmo Mastroserio$^{3}$, \newauthor Michiel van der Klis$^{4}$, Matteo Lucchini$^{4}$, Jake Mitchell$^{1,5}$ \\ 
$^{1}$School of Mathematics, Statistics and Physics, Newcastle University, Herschel Building, Newcastle-upon-Tyne, NE1 7RU, UK\\
$^{2}$California Institute of Technology, Pasadena, CA 91125, USA\\
$^{3}$Dipartimento di Fisica, Universit\`a degli Studi di Milano, Via Celoria 16, I-20133 Milano, Italy\\
$^{4}$Anton Pannekoek Institute for Astronomy, University of Amsterdam, Science Park 904, 1098 XH Amsterdam, The Netherlands\\
$^{5}$Leibniz-Institut f\"ur Astrophysik Potsdam (AIP), An der Sternwarte 16, 14482 Potsdam, Germany\\
}

\date{Accepted 2025 December 13. Received 2025 November 22; in original form 2025 January 20}

\pubyear{\the\year{}}

\begin{document}
\label{firstpage}
\pagerange{\pageref{firstpage}--\pageref{lastpage}}
\maketitle
\begin{abstract}  
We fit X-ray reverberation models to \textit{Rossi X-ray Timing Explorer} data from the X-ray binary Cygnus X-1 in its hard state to yield estimates for the black hole mass and the distance to the system. The rapid variability observed in the X-ray signal from accreting black holes provides a powerful diagnostic to indirectly map the ultra-compact region in the vicinity of the black hole horizon. X-ray reverberation mapping exploits the light crossing delay between X-rays that reach us directly from the hard X-ray emitting `corona’, and those that first reflect off the accretion disc. Here we build upon a previous reverberation mass measurement of Cygnus X-1 that used the \textsc{reltrans} software package. Our new analysis enhances signal to noise with an improved treatment of the statistics, and  implements new \textsc{reltrans} models that are sensitive to distance. The reduced uncertainties uncover evidence of mass accretion rate variability in the inner region of the disc that propagates towards the corona\rev{, which we approximate as a point-like `lamppost' source.
Our best fitting model returns a mass of $M=15\pm 4 ~M_\odot$ and a distance of $D=3.4_{-1.2}^{+1.6}$ kpc (90 per cent uncertainties), which are consistent with the most recent dynamical and parallax measurements respectively.}
    
\end{abstract}

\section{Introduction}
\label{sec:intro}
Black hole \xray\ binary (BHXB) systems are powerful \xray\ sources, powered by accretion of matter from a companion star onto the black hole. The black hole is thought to accrete via a thermally radiating disc that is geometrically thin and optically thick \citep{SS:1973,Novikov:1973}. \revnew{Part of the accretion flow is a component colloquially referred to as the \textit{corona} \citep{Thorne:1975}; a region of hot, optically thin plasma. Soft thermal X-rays produced by the disk ($\sim 0.1-1$ keV) are Compton up-scattered by hot electrons in the corona.} 
\revnew{The up-scattered photons emergent from the corona form the high energy X-ray spectrum referred to here as the \textit{continuum emission} which}
can be modelled with a cut-off power law. 
The power law index and cut-off energy are jointly related to the optical depth and electron temperature ($kT_e$) of the corona.

The exact nature of the corona is still a subject of ongoing research \citep[see e.g.][for a recent review]{Poutanen:2018}. Suggested geometries include a patchy layer above the disc \citep{Galeev:1979}, the base of a vertically extended out-flowing jet \citep{Markoff:2005,Kylafis:2008}, and a large scale-height inner accretion flow inside of a truncated disc \citep[the \textit{truncated disc} model;][]{Eardley:1975,Done:2007}. Recent measurements of X-ray polarisation aligning with the large-scale radio jet indicate that the corona is extended in the plane perpendicular to the jet, disfavouring the vertically extended models \citep{Henric:2022,Veledina:2023,Ingram:2024}. However, for mathematical convenience, the corona is often approximated as a point-like source above the black hole \citep[the \textit{lamppost model};][]{Matt:1992,Miniutti:2004}.

Irrespective of the coronal geometry, the continuum photons that travel directly to us dominate the hard part of the observed \xray\ spectrum; whereas the photons following trajectories that intercept the disc are re-processed and re-emitted, in a process referred to as \textit{reflection}. The rest-frame reflection spectrum exhibits features such as an iron K$\alpha$ line at $\sim 6.4$ keV, a broad Compton hump peaking in the range $\sim 20-30$ keV, and a soft excess \citep{George:1991,Ross:2005,Fabian:2010,Garcia:2010}. Owing to gravitational distortions and rapid orbital motion of the inner accretion disc we see a smeared reflection spectrum \citep{Fabian:1989}. The same reflection features are seen in the X-ray spectrum of active galactic nuclei (AGNs), which are also thought to accrete via a disc and corona.

The coronal emission is highly variable, which causes the subsequent reflection to also vary after a light-crossing time, often  
\rev{referred to as \textit{reverberation lags}: reflection-dominated energy bands lag continuum-dominated bands}. 
They can be accessed observationally by calculating the cross spectrum (the Fourier transform of the cross-correlation function) between light curves extracted from different energy bands \citep{vanderklis:1987}. The cross spectrum is a complex quantity that depends on Fourier frequency $\nu$, where higher Fourier frequencies correspond to more rapid variability. The phase lag $\phi(\nu)$ between the bands is equal to the argument of the cross spectrum, and the time lag is $t_{\rm lag}(\nu) = \phi(\nu)/(2\pi\nu)$. A lag vs energy spectrum can be constructed by calculating time lags in a given Fourier frequency range between several different `subject' energy bands and one common `reference' band. A reverberation signal manifests as reflection features (i.e. soft excess, iron line, Compton hump) in the lag vs energy spectrum. Clear iron line reverberation features have been detected for $\gtrsim 25$ AGNs \citep{Kara:2016} and at least one BHXB \citep{Kara:2019}. So called `soft lags' between the 0.5-1 keV band that includes the soft excess and the continuum-dominated 2-5 keV band have been detected for many observations of many AGNs and BHXBs \citep{Fabian:2009,DeMarco:2013,Wang:2022}, although the association of this soft lag with reverberation is less certain than for the iron line feature \citep[e.g.][]{Uttley:2025}.

Modelling of the reverberation features in the lag vs energy spectrum enables a measurement of black hole mass, since the reverberation lag is proportional to the light-crossing timescale of the black hole. Several codes now exist to enable such modelling \citep[e.g.][]{Cackett:2014,CaballeroGarcia:2018}. In this paper we use models from the X-ray reverberation mapping package \reltrans \citep{Ingram:2019a}. The \reltrans models employ a lamppost geometry, whereby the corona is a point-like source a height $h$ above the black hole, and utilise grids from the \textsc{xillver} model to calculate the restframe reflection spectrum \citep{Garcia:2013a}.

Reverberation features, however, are only seen at high Fourier frequencies (\rev{$\gtrsim 100 ~M_\odot/M$ Hz}, where $M$ is black hole mass), corresponding to the fastest variability timescales. At lower frequencies, the lag instead increases log-linearly with energy \citep{Kotov:2001}. These lags are often referred to as \textit{hard lags}, since hard photons always lag soft photons, or \textit{continuum lags}, since there are no reverberation features. The magnitude of the continuum lag decreases with frequency \citep[e.g.][]{Nowak:1999} such that at low frequencies it completely dominates over the reverberation lag, whereas at high frequencies it becomes small enough for the reverberation lag to dominate (see reviews by \citealt{Uttley:2014,Bambi:2021}). The higher Fourier frequencies can therefore be fit with a pure reverberation model, but this leaves a degeneracy between the source height and the black hole mass \citep{Cackett:2014,Ingram:2019a}. The degeneracy can be broken either by simultaneous modelling of many epochs between which the coronal geometry changes \citep{Alston:2020}, or by simultaneously modelling a range of Fourier frequencies \citep{Mastroserio:2018}. The latter is far less expensive in terms of observing time, but requires the continuum lags to be `modelled out' with a reasonable model.

The coronal variability is likely due to propagating fluctuations in the mass accretion rate \citep{Lyubarskii:1997, Kotov:2001, Arevalo:2006, Rapisarda:2016}. This process can drive the continuum lags via one, or a combination of, two mechanisms. The first is stratification of the corona: fluctuations first modulate the soft X-rays emitted from the outer corona, before propagating to the hard X-ray emitting inner corona \citep[e.g.][]{Arevalo:2006,Mahmoud:2018,Kawamura:2022}. The second is variable heating and cooling: a perturbation in the disc first cools the corona via an increase in seed photons, before propagating into the corona to heat it \citep{Uttley:2025}. Such variable heating and cooling leads to variations in the power law index $\Gamma$ of the Comptonised spectrum, so-called \textit{spectral pivoting}, which naturally reproduces the energy dependence of the continuum lags \citep{Kotov:2001,Koerding:2004}. The \reltrans models implement spectral pivoting in an \textit{ad hoc} manner to model out the continuum lags with minimal computational expense. The influence of this spectral pivoting on the resulting reflection spectrum is included using first-order Taylor expansions \citep{Mastroserio:2018,Mastroserio:2021}.

In an early successful demonstration of the \reltrans package, \citet[][hereon M19]{Mastroserio:2019} used X-ray reverberation mapping to measure the mass of the black hole in Cygnus X-1 (hereafter Cyg X-1). Instead of fitting to the time lags, M19 employed joint fits to the time-averaged spectrum and the cross spectrum\footnote{M19 actually fit to the `complex covariance', which differs from the cross spectrum only by an arbitrary normalization factor.} (real and imaginary parts) in 10 frequency ranges. Fitting to the cross spectrum itself has two main advantages. First, it is statistically favourable because different model components can be added together \citep{Mastroserio:2018}. Second, it also considers the \textit{amplitude} of variability in addition to the lags. This provides important extra constraints, since fast variability in the reflected signal is washed out by destructive interference between rays reflected from different parts of the disc \citep{Revnivtsev:1999}. M19 found $M = 26_{-9}^{+10}~M_\odot$, which agrees remarkably well with the
dynamical mass measurement of $M = 21\pm 2~M_\odot$ \rev{that was published shortly after} \citep{MillerJones:2021}.

Here, we revisit the M19 analysis, using the same RXTE dataset. One motivation for this is to address the low $\chi^2$ of the M19 model fits that was caused by the uncertainties in the cross-spectrum data being overestimated. We improve on the the M19 analysis by employing revised error formulae \citep{Ingram:2019b}. The main motivation, however, is to test the new features that have been added to the \reltrans models since 2019. The most prominent change is that the models are now sensitive to the distance $D$ to the source \citep{Ingram:2022}. This is because the shape of the reflection spectrum is sensitive to the ionization state of the disc. Therefore, put simply, if the disc appears to be highly ionised but the observed flux is low, then $D$ must be large. Such an inference of $D$ was not previously possible because the ionization state of the disc does not only depend on the illuminating flux, but also on the disc density \citep{Garcia:2016}, which for many years was hardwired in reflection models to a value of $n_e=10^{15}$ electrons per cm$^3$. New \textsc{xillver} grids now enable $n_e$ to be fit as a free parameter, meaning that it is now possible to simultaneously infer both the illuminating flux and the disc density from reflection spectroscopy \citep{Jiang:2019}. In this paper, we therefore use new \reltrans models that employ the density-dependent \textsc{xillver} grids to fit for the distance to Cyg X-1 as well as the mass.

Cyg X-1 is an ideal source for this analysis because the distance and mass \rev{have been measured via other techniques,} 
thus providing a means to test the reverberation model and constrain geometrical parameters. \rev{The distance is robustly constrained by parallax measurements to be $D = 2.2\pm 0.2$ kpc \citep{MillerJones:2021}. The mass is more model dependent. 
Originally \cite{MillerJones:2021} inferred $M = 21 \pm 2~M_\odot$, but recently \cite{Ramachandran:2025} used the same data and more sophisticated modelling of the companion star to infer a lower mass. The new analysis returns $M=17.5^{+2}_{-1}~M_\odot$ when adopting the best fitting inclination of $i=27.5^\circ$ from previous analyses. However, when a range of possible inclination angles are considered, uncertainties increase. In this paper, we compare our results to both the \cite{MillerJones:2021} and \cite{Ramachandran:2025} values.} We use the same dataset as M19 to provide a solid starting point for our analysis. \rev{We aim to use this dataset to test if the model can infer reasonable distance and black hole mass values for Cyg X-1, with the future goal of using it on black hole X-ray binaries for which these parameters are not currently known.}

The structure of this paper is as follows. In Section \ref{sec:data} we describe the observations used and the generation of the cross spectra and their uncertainties. Section \ref{sec:models} details how we construct and fit our reflection models. In Section \ref{sec:results} we present the results of our modelling; and in Sections \ref{sec:discussion} and \ref{sec:conclusions} we discuss our findings and present our conclusions.  

\section{Data}
\label{sec:data}
\subsection{Data reduction}

We consider the same archival RXTE observations taken in 1996 that were analysed by M19. These are the final five of seven observations from the proposal P10238. We stack these five observations together, since their spectra are very similar in shape to one another. We use only Proportional Counter Array (PCA) data. For fits to the time-averaged spectrum, we use `standard 2' data. The timing data are in `generic binned' mode, which has $dt=1/64$ s time resolution in 64 energy channels covering the whole PCA energy band. All five proportional counter units (PCUs) were switched on for the entirety of all the five observations we stack over. Details of our data reduction procedure can be found in \citet{Mastroserio:2018}.

\subsection{Calculation of cross spectra}

To calculate cross spectra, we first extract 64 subject band light curves, $s(E_n,t_k)$, where $s$ is the count rate summed over all five PCUs in the $n^{\rm th}$ energy channel and the $k^{\rm th}$ time bin. We then construct a reference band light curve, $r(t_k)$, by summing over a subset $n_{\rm min}$ to $n_{\rm max}$ of the 64 subject band light curves
\begin{equation}
    r(t_k) = \sum_{n=n_{\rm min}}^{n_{\rm max}} s(E_n,t_k).
\end{equation}
\rev{We employ different choices for the reference band in different parts of our analysis, first choosing a narrow 2.8-3.7 keV band for consistency with M19, before moving to a broader 3.1-24.7 keV band to maximise signal to noise.}
For each energy channel, we calculate a cross spectrum \citep{Ingram:2019b}
\begin{equation}
    G(E_n,\nu_j) = \langle S(E_n,\nu_j) R^*(\nu_j) \rangle - \mathcal{N}(E_n,\nu_j).
\end{equation}
Here, upper case letters represent the Fourier transform of the corresponding lower case letter, and $\nu_j$ is Fourier frequency. Throughout this paper, we employ absolute rms normalization for Fourier transforms, meaning that the integral of the power spectrum over all positive frequencies is equal to the variance of the corresponding time series \citep[e.g.][]{Uttley:2014}. The angle brackets represent ensemble averaging over segments of length 
\rev{$T_{\rm seg} = 2^{15} dt = 512$ s}
and over a range of discrete Fourier frequencies \citep[see e.g.][]{vanderKlis:1989} centred on $\nu_j$. Throughout the paper, we consider the same 10 frequency ranges derived from a geometrical re-binning scheme with re-binning constant $c_0=2.5$ \rev{and a minimum binning of 6} \citep[see][]{Ingram:2012}. The lowest frequency range is $9.8\times10^{-4} - 4.9 \times 10^{-3}$ Hz and the highest is $12 - 32$ Hz. The term $\mathcal{N}(E_n,\nu)$ represents the Poisson noise to subtract from the cross spectrum \citep{Ingram:2019b}. We describe our treatment of Poisson noise in Appendix \ref{sec:pois}.
\begin{figure*}
    \centering
    \includegraphics[width=\linewidth,trim=0 0.8in 0 0,clip=true]{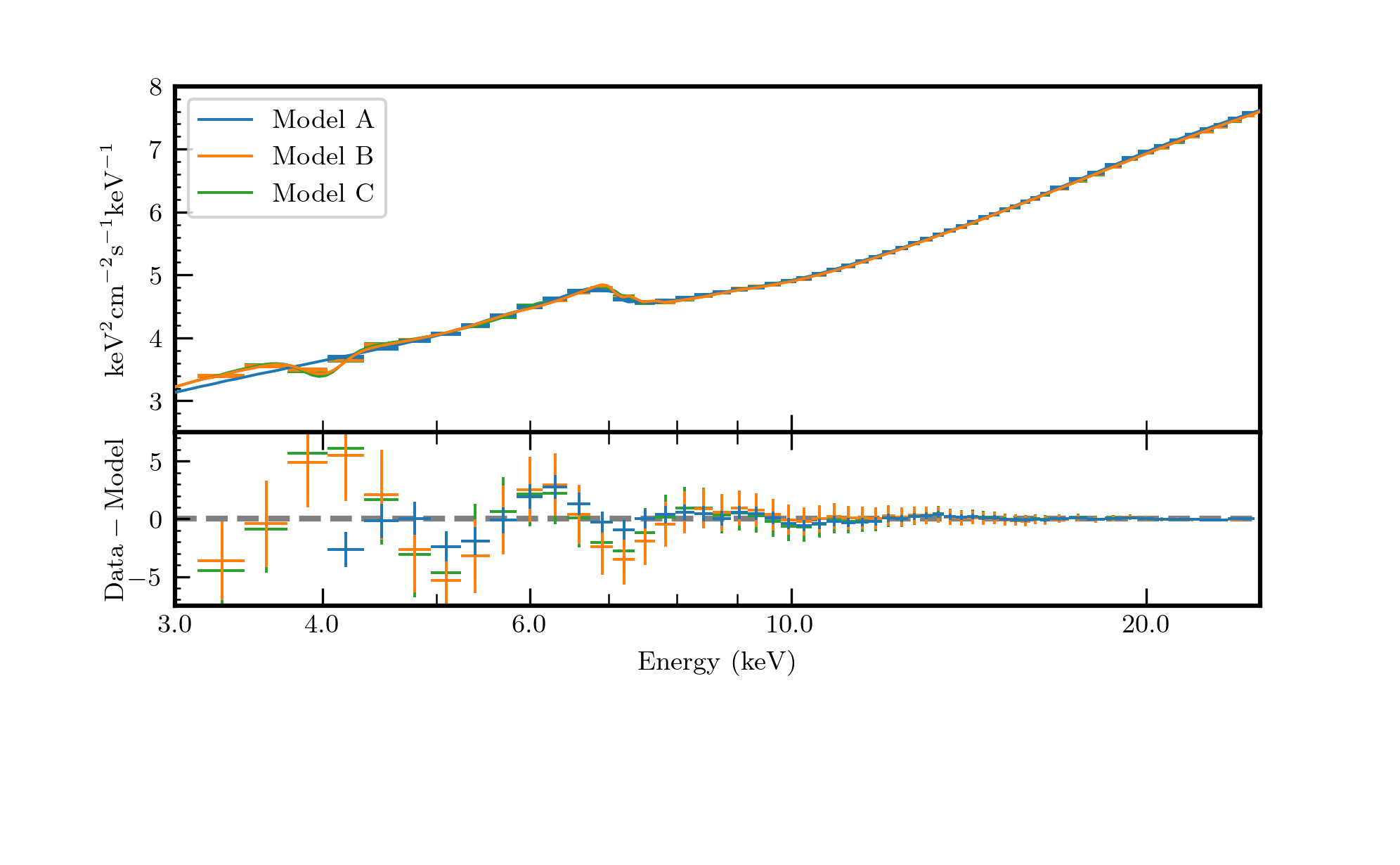}
    \vspace{-20pt}
    \caption[TAspectrum]{Top Panel: Unfolded time-average spectrum using the \reltransdcp (blue and orange) and \rtdist (green) reflection models. Bottom panel: residuals of the time averaged spectrum. These models were obtained via fitting the time-averaged and cross spectra simultaneously.}
    \label{fig:TA}
\end{figure*}

\subsection{Uncertainty estimates}
\label{sec:error_estimate}

In this paper, we fit energy-dependent models to the real and imaginary parts of the cross spectrum in 10 frequency bands. This extracts the same information as fitting simultaneously to the time lag and variability amplitude as a function of energy in the same frequency ranges, but with several statistical advantages \citep{Mastroserio:2018}.

We employ two estimates for the 1 $\sigma$ statistical uncertainties on the cross spectra. The first uses the analytical formulae of \citet[][hereafter the BP formulae]{BP:2000}. The BP formulae for the uncertainty on the real and imaginary parts of the cross spectrum can be found in Equation 13 of \citet{Ingram:2019b}. These formulae are appropriate for fitting a model for the frequency dependence of a single cross spectrum (i.e. only one subject band and one reference band), and give the same result as estimating errors from the standard deviation around the ensemble-averaged values of the cross spectrum (see e.g. Fig. 4 of \citealt{Ingram:2019b}). This error estimate was used by M19, but it was found to produce \revnew{errors} that are too large. We use the BP formulae here as an initial consistency check with M19.

\rev{The over-estimation of the \revnew{errors} was not well understood at the time of the M19 analysis, but was subsequently cleared up by \cite{Ingram:2019b}.} The second error estimate we employ \rev{therefore} uses Equation (18) of \citet{Ingram:2019b}, for which the uncertainty on the real and imaginary parts of the cross spectrum are equal to one another. This formula (hereafter the \rev{I19} formula) is appropriate for fitting a model to the cross spectrum as a function of energy for a given frequency range, as we do in this paper. The different formula is required because the reference band is the same for each subject band considered (and because variability is strongly correlated between energy channels). Using the BP formula accounts for the same statistical uncertainty on the reference band many times, once for each subject band. \citet{Ingram:2019b} showed that this was the source of the large \revnew{errors} implemented by M19. The \rev{I19} formula instead treats the statistical uncertainty on the reference band as a systematic uncertainty in the variability amplitude (which is not of primary physical interest to us). We therefore switch to the \rev{I19} formula after verifying consistency with M19.

The two error estimates converge in the Poisson noise dominated regime, which our data reach at high frequencies. For example in the frequency band above $12$ Hz ($12-32$ Hz), the \rev{I19} and BP formulae produce cross spectral errors that at within $0.7$ per cent of each other. Whereas in the lowest cross spectrum frequency band ($\sim 0.001-0.005$ Hz), the \rev{I19} error formula produces errors  $\sim80$ per cent smaller than the BP errors (see e.g. Fig. 5 of \citealt{Ingram:2019b}).

\section{Reverberation Modelling}
\label{sec:models}
Using \xspec version 12.13.1 \citep{Arnaud:1996}, we fit models from the \reltrans package \citep[version \rev{2.2}:][]{Mastroserio:2021} simultaneously to the time-averaged spectrum and real and imaginary parts of the cross spectrum in 10 frequency bands \rev{(i.e. a total of 21 spectra)}. \rev{Unless stated otherwise, we apply $0.4$ per cent systematic errors to all spectra (time-averaged and cross spectra) to account for uncertainties around the spectral calibration of the PCA. We note that the real and imaginary parts of the cross-spectrum are proportional to the time-averaged count rate, so any systematic error applied to the time-averaged spectrum must also be applied to the cross spectra (see Appendix \ref{sec:sys} for the mathematical details).}
\newline
\subsection{Model setup}

\rev{We consider two \reltrans model flavours: \reltransdcp and \rtdist. The most obvious difference between the two is that \rtdist takes the distance as a model parameter, whereas \reltransdcp does not. However, the distance can be \textit{inferred} from the best fitting \reltransdcp parameters, and here we present a post-processing code to return a distance measurement from a \reltransdcp fit (see Section \ref{sec:MandD}). There are other, more subtle, physical differences between the two models, and thus here we conduct fits with both. However, before conducting the new fits, we first aim to reproduce the best fit of M19 as a consistency check (Model A), for which we use \reltransdcp (since this is the closest model in the current package to the models employed by M19). We then make} several improvements to the M19 analysis to yield new fits using \reltransdcp (Model B) and \rtdist (Model C).

Throughout, we include a radial dependence of the ionization parameter $\xi(r)$ and disc  electron number density $n_e(r)$ by splitting the disc into 20 radial zones (\texttt{ION\_ZONES}=20; which we have verified to be adequate to reach convergence). The ionization parameter is defined as
\begin{equation}
    \xi(r) = \frac{4\pi F_x(r)}{n_e(r)},
\end{equation}
where $F_x(r)$ is the illuminating X-ray ($0.1-1000$ keV) flux, which the model calculates self-consistently. We assume the electron density profile of a zone A Shakura-Sunyaev accretion disc \citep{SS:1973}, an assumption that yielded the best fitting model of M19. \rev{The \reltrans\ models} account for the angular dependence of the emergent reflection spectrum \citep{Garcia:2014a} by integrating over \rev{a fixed number of} 
angular zones (\rev{set by the environment variable} \texttt{MU\_ZONES}). \rev{Here, we use only one angular zone (\texttt{MU\_ZONES}=1) which is adequate for the low inclination angle of Cyg X-1 \citep{Ingram:2019a}.}
Throughout, we fix the BH spin to $a=0.998$ and allow the disc inner radius $r_{\rm in}$ to be a free parameter. This enables us to explore the largest possible range of $r_{\rm in}$ without it becoming smaller than the ISCO. Although these reflection models are sensitive to the BH spin itself, through the null geodesics and energy shifts, they are much more sensitive to $r_{\rm in}$. We freeze the disc outer radius to $r_{\rm out}=2 \times 10^4~R_g$, where $R_g=GM/c^2$ is the gravitational radius.
We account for line-of-sight absorption using the \xspec model \tbabs with the relative abundances of \cite{Wilms:2000}.

All models we use approximate the corona as a stationary lamppost at height $h$ aligned along the black hole spin-axis; and the accretion disc as Keplerian with a constant aspect ratio $z/r$ where $z$ is the scale height. The aspect ratio in \reltransdcp is hard-wired to zero, whereas in \rtdist\ $z/r$ is a model parameter. Preliminary analysis indicated that the \rtdist\ fits were not sensitive to $z/r$, so we fix it to $z/r=0$ for all fits presented here. 
\begin{figure}
    \centering
    \includegraphics[width=\linewidth]{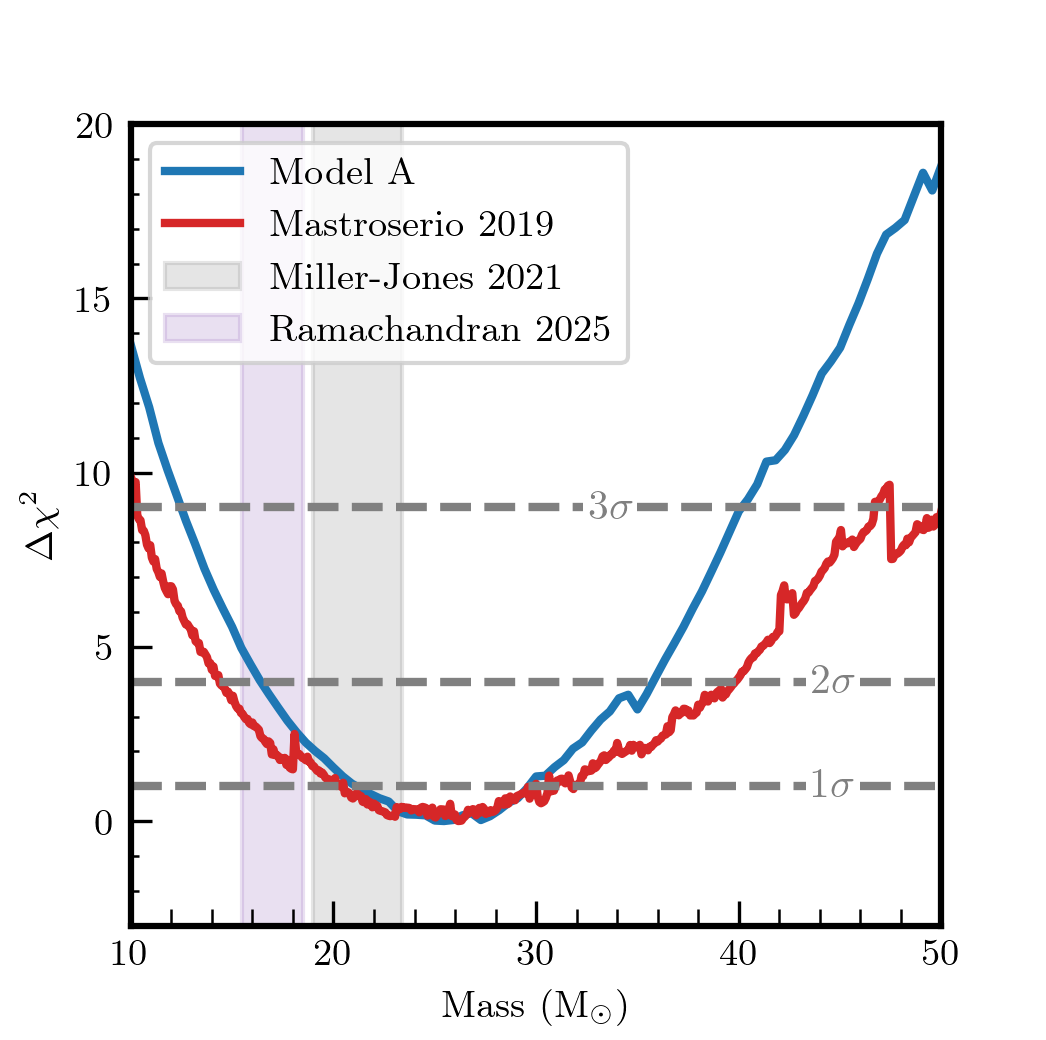}
    \vspace{-10pt}
    \caption[Steppar analysis]{Comparison of the $3\sigma$ mass, $M$, interval obtained in M19 (red) and our Model A (blue). The $3\sigma$ confidence interval obtained in M19 constrains $M\sim10-50$ M$_{\odot}$; while the same confidence interval obtained here constrains $M\sim15-40$ M$_{\odot}$. Both contours are in agreement with dynamical mass measurements indicated by the grey and purple shaded regions.}
    \label{fig:1d_steppars}
\end{figure}
\subsection{Model A: reproduction of M19}
\label{sec:Gullo_data}
\begin{figure*}
    \centering
    \includegraphics[width=\linewidth]{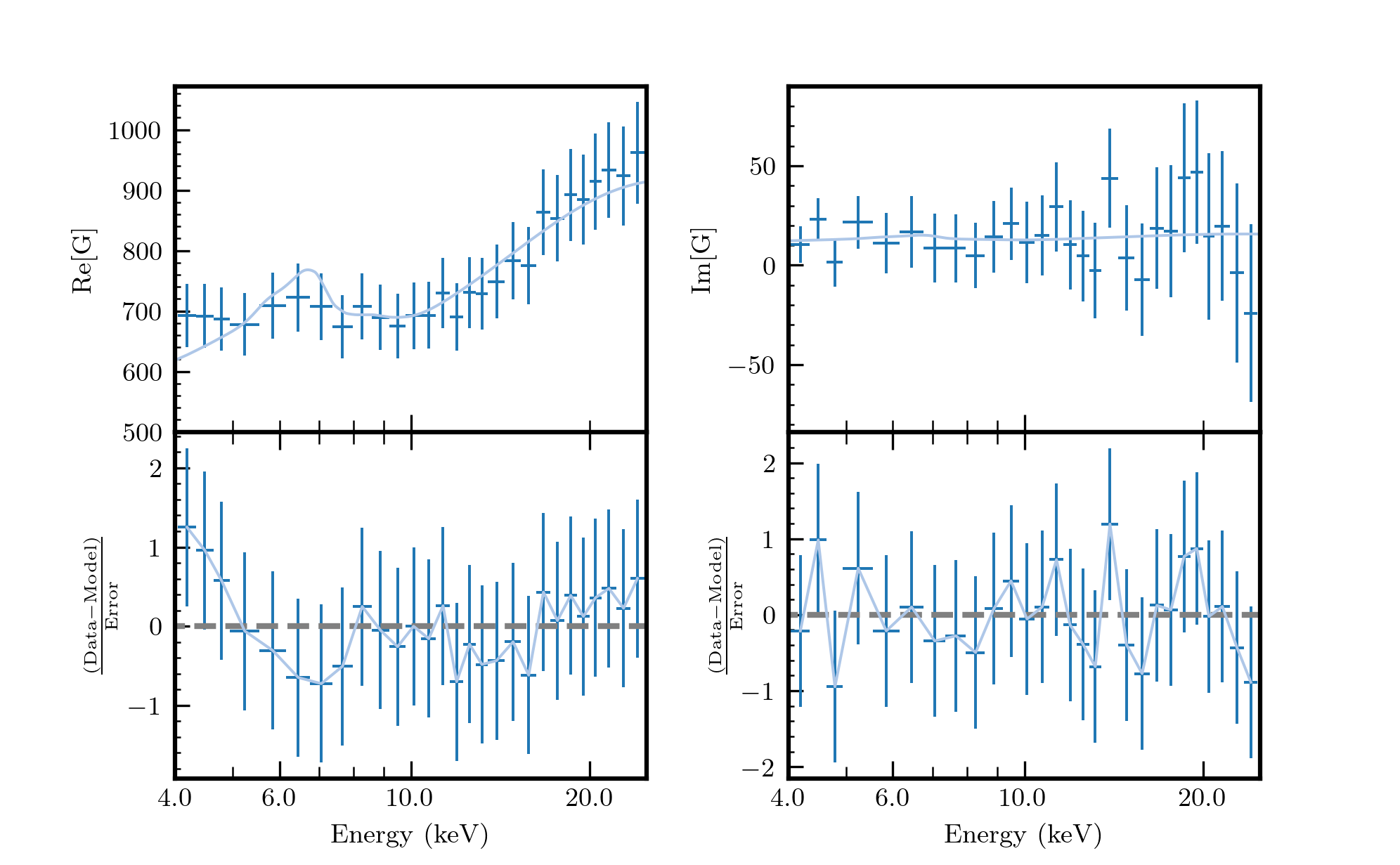}
    \vspace{-20pt}
    \caption[narrow ref cross spectrum]{Unfolded cross spectrum (top) and contributions to the overall $\chi^2$ (bottom) in a single frequency band ($0.001 - 0.005$ Hz), with uncertainties calculated using \citet{BP:2000} formulae (Model A). Real and imaginary parts are plotted on the left and right respectively. We see that the error bars are overestimated.}
    \label{fig:BP_cs}
\end{figure*}

We start by setting up a fit designed to reproduce the M19 results as closely as possible. It is not possible to exactly reproduce the M19 fit due to several bug fixes and extra physics in the latest version of the model. These improvements are described in detail in \citet{Mastroserio:2020} and \citet{Mastroserio:2021}. Another key difference is that the \reltrans\ model used by M19 implements a power law with a high energy \rev{exponential} cut-off for the continuum, whereas here we use \reltransdcp, which implements the thermal Comptonisation model \texttt{nthcomp} \citep{Zdziarski:1996} for the continuum. We are also now fitting to the cross spectrum itself instead of the ‘complex covariance’ (as in M19), but these two statistics differ only in arbitrary normalization. \rev{It is an important first step of our analysis to verify that the described changes to the model since its original version do not cause large discrepancies with the results of M19.}

We use the following model: 
\begin{equation}
    \texttt{TBabs} \times \reltransdcp.
    \label{eq:BP_dcp}
\end{equation}
We set the disc density to make the model as similar as possible to the one used in M19. An exact match is not possible, since in old versions of \reltrans models, the density was always hardwired to $n_e = 10^{15}$ cm$^{-3}$ for the calculation of the restframe reflection spectrum, whereas in the new models the disc density is a function of radius, $n_e(r)$, and the model parameter is the minimum of that function. We fix this minimum density to $n_e = 10^{15}$ cm$^{-3}$. Following M19, we use a reference band of $2.8-3.7$ keV, calculate errors with the BP formula,
only consider the $4-25$ keV energy range in our fits\rev{, and apply a $0.1$ per cent systematic error to the time-averaged spectrum only}\footnote{\rev{We do this to follow the M19 analysis exactly. No systematic was applied to the cross spectra in M19, since the cross spectral uncertainties were already large (overestimated, as it was subsequently realised). A smaller systematic could be applied to the spectrum because the $3-4$ keV region was ignored, which is the most problematic in terms of calibration.}}. 

We refer to this as Model A. Fig. \ref{fig:TA} (blue) shows the best-fitting time averaged spectrum, and Table \ref{tab:params} (first column) shows the best fitting parameters. The fit statistic for our Model A is $\chi^2/{\rm dof} = 492/563$, which is larger than the $\chi^2$ for M19 best fitting model (their Model 3\rev{: $\chi^2/{\rm dof} = 426/563$), but is still a reduced $\chi^2$ smaller than unity.}

Fig. \ref{fig:1d_steppars} shows $\Delta\chi^2$ versus mass for Model A (blue), alongside that of M19 (red). \rev{We see that the two mass contours are consistent with one another.} The low $\chi^2$ is due to the use of the BP error formulae to calculate the uncertainties on the real and imaginary parts of the cross spectrum \citep{Ingram:2019b}. This can be seen in Fig. \ref{fig:BP_cs}, which presents the real and imaginary parts of the cross spectrum in one representative frequency range. We see that the error bars are clearly larger than the dispersion of the data around the model, indicating that they are over estimated.

\begin{figure}
    \centering
    \includegraphics[trim={0 1in 0 0},clip,width=\linewidth]{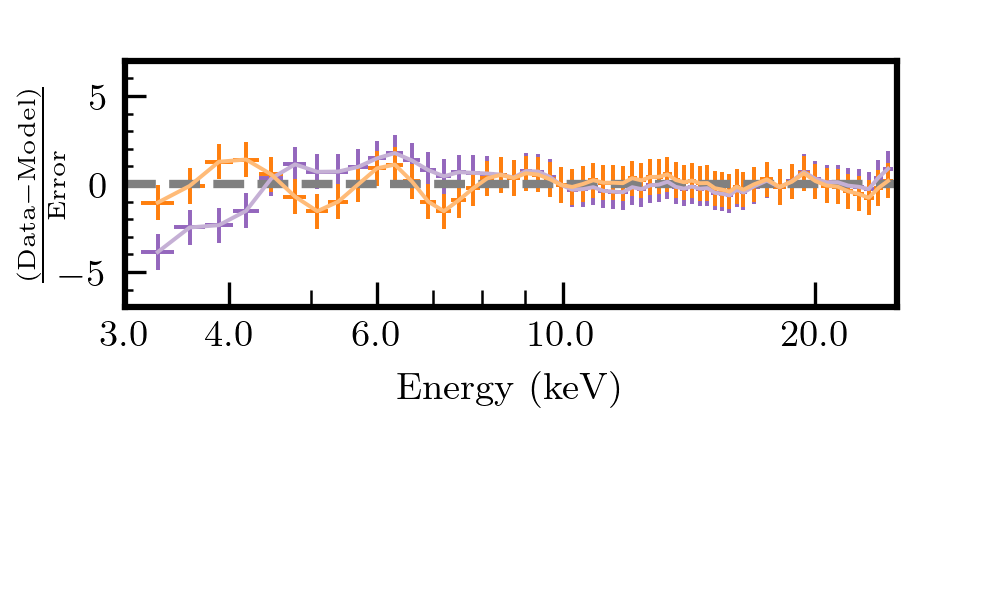}
    \includegraphics[trim={0 0.55in 0 0},clip,width=\linewidth]{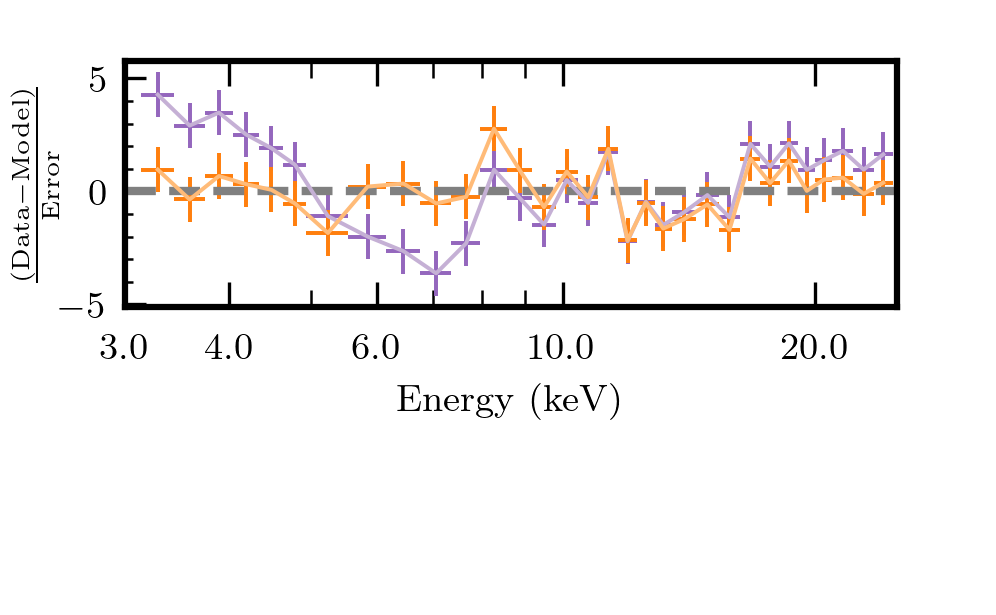}
    \vspace{-20pt}
    \caption[wide ref no extras cross spectrum]{Contributions to $\chi^2$ obtained by two models for the time-averaged spectrum (\textit{top}) and the real part of the cross spectrum in a single frequency band (\textit{bottom}). Orange points are for our Model B, whereas purple points are for \texttt{TBabs} $\times$ \reltransdcp only (i.e. with no disk or calibration components). Although we only show the real part of the cross spectrum for a single frequency band ($\nu \sim 0.001 - 0.005$ Hz), similarly structured residuals at low energies are present at all frequencies.}
    \label{fig:comparison}
\end{figure}

\begin{figure*}
    \centering
    \includegraphics[trim={0cm 1.175in 0cm 0cm},clip,width=\linewidth]{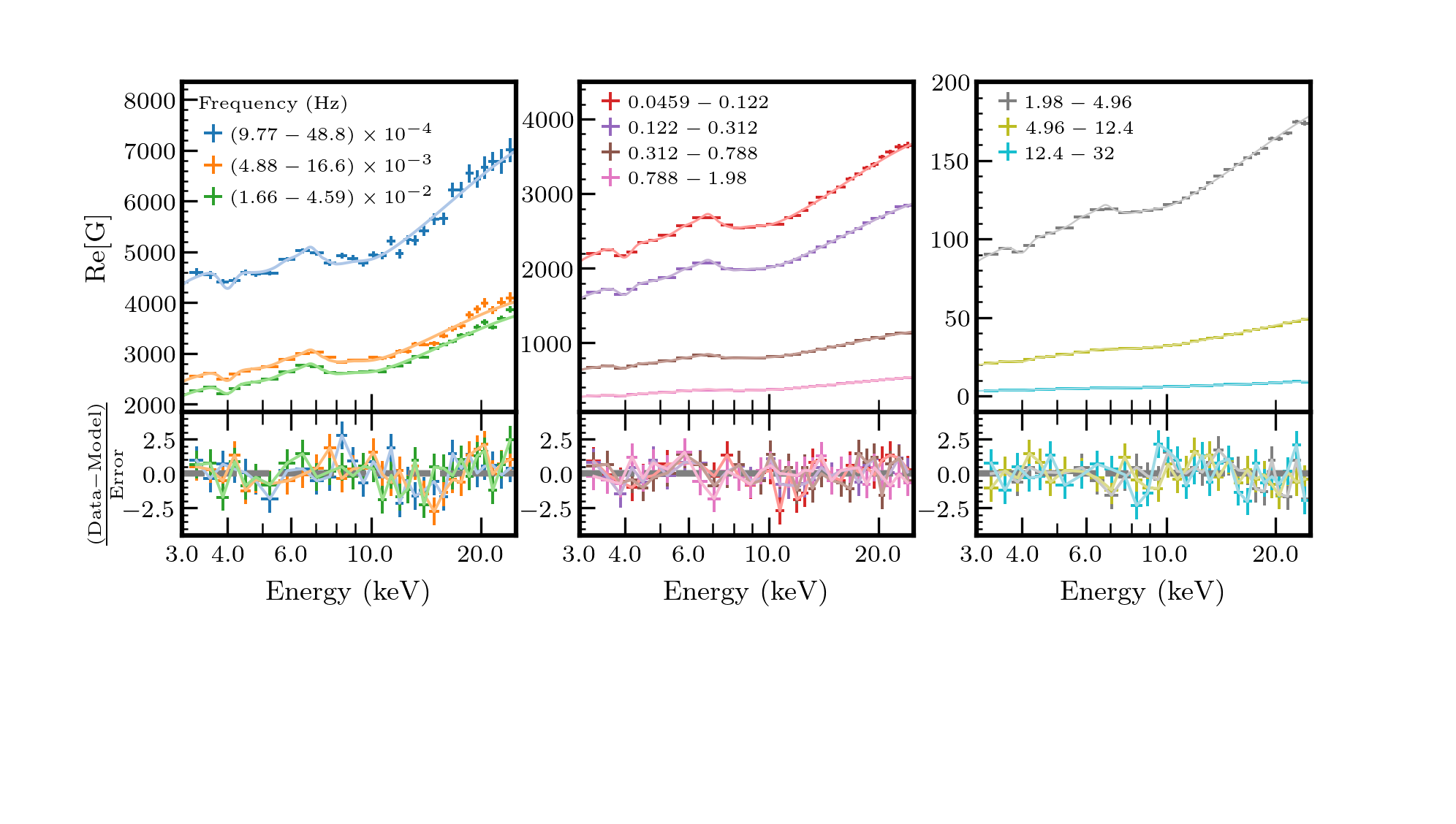}
    \includegraphics[trim={0cm 1in 0cm 0.5cm},clip,width=\linewidth]{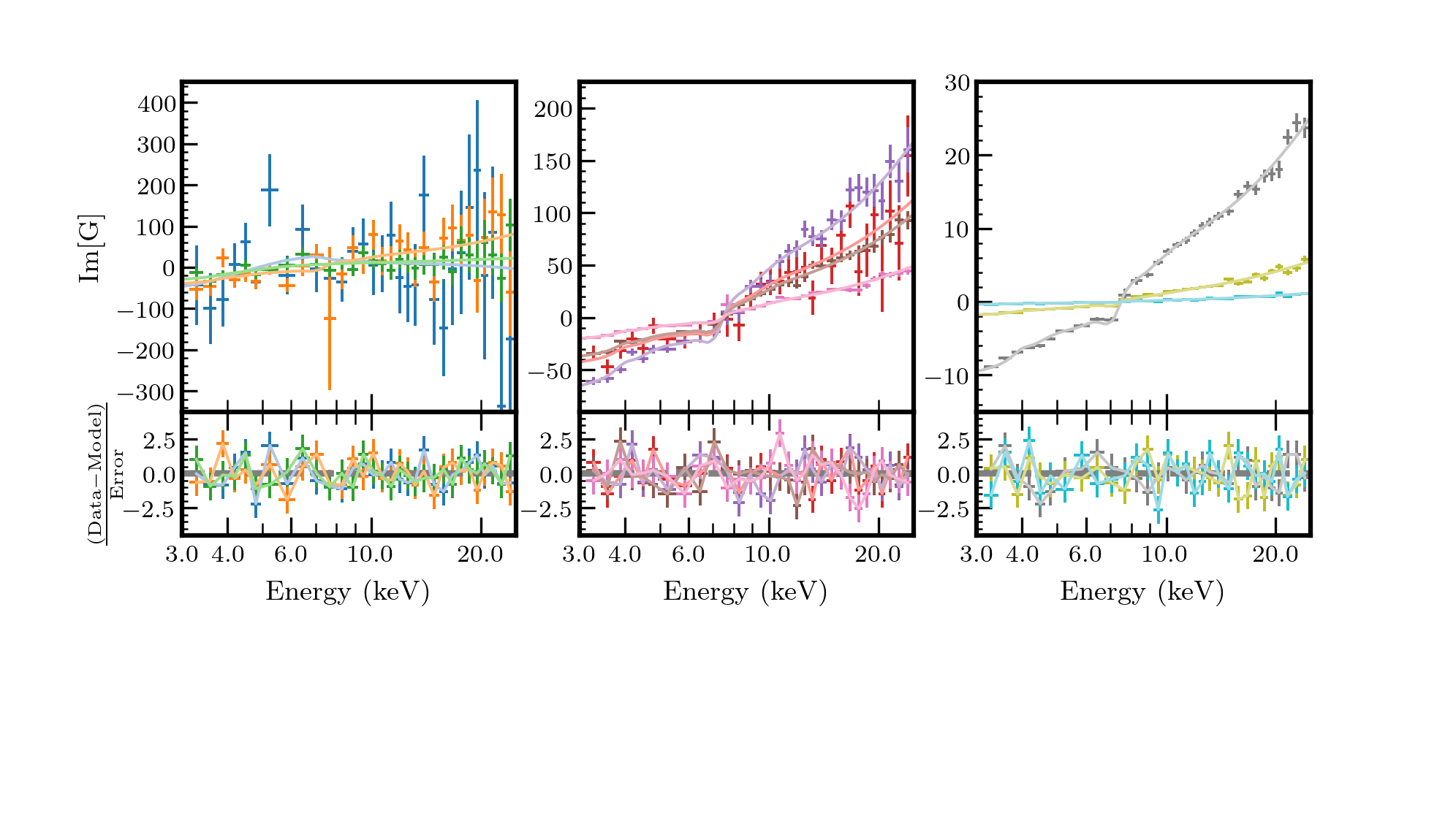}
    \vspace{-15pt}
    \caption[wide ref dcp + extras cross spectrum]{Model B: Unfolded real (top panels) and imaginary (bottom panels) parts of the cross spectrum in all 10 frequency bands. The bottom axis of each panel show the contributions to $\chi^2$.}
    \label{fig:AIDCP}
\end{figure*}

\subsection{Model B: free disc density and improved uncertainty estimates}
\label{sec:modelB}

We now improve on M19 whilst continuing to use \reltransdcp. In this iteration we make several changes: 1) we allow the minimum disc density $n_e$ to be a free parameter. 2) We extend the fitting range to 3-25 keV from 4-25 keV to increase the sensitivity of the model to $n_e$. 3) We use the \rev{I19} error formulae for the cross spectra to address the problem of large \revnew{errors}, implemented in the original M19 analysis. 4) We increase signal to noise by extending the energy range of the reference band to 3.1-24.7 keV.

\begin{figure*}
    \centering
    \includegraphics[trim={0cm 0cm 0cm 0cm},clip,width=\linewidth]{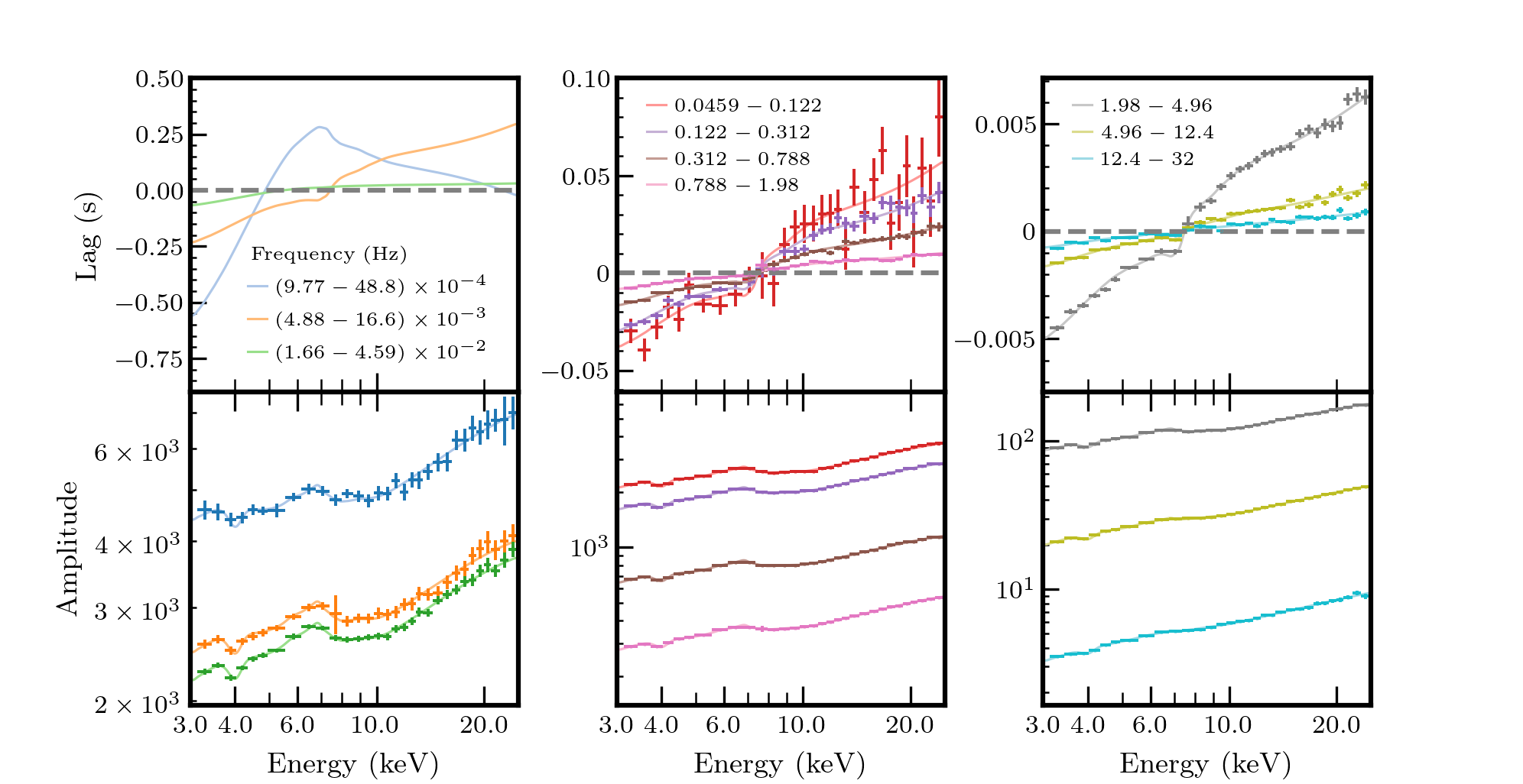}
    \caption{Model B: Lag-energy spectrum (top) and variability amplitude (bottom panels) for different Fourier frequency bands. The points are calculated from the unfolded cross spectra and lines are calculated using parameters of Model B. We only show the model lines for the smallest frequency bands, the data in these frequency bands are not shown due to their large \revnew{errors}. The units of amplitude are keV$\,$cm$^{-2}\,$s$^{-1}$.}
    \label{fig:dcp_lag_amp}
\end{figure*}

\begin{figure*}
    \centering
    \includegraphics[trim={0cm 1.175in 0cm 0cm},clip,width=\linewidth]{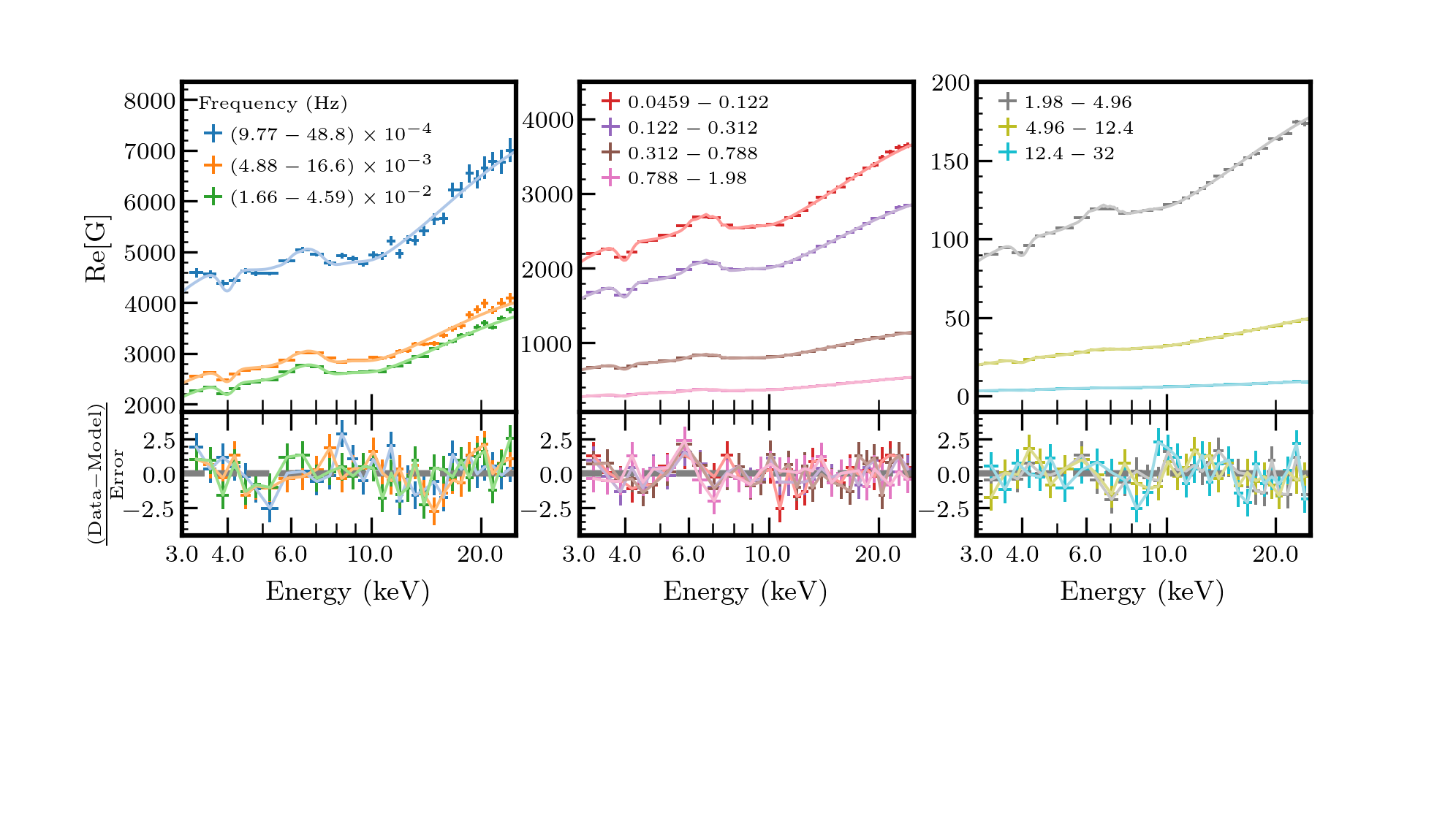}
    \includegraphics[trim={0cm 1in 0cm 0.5cm},clip,width=\linewidth]{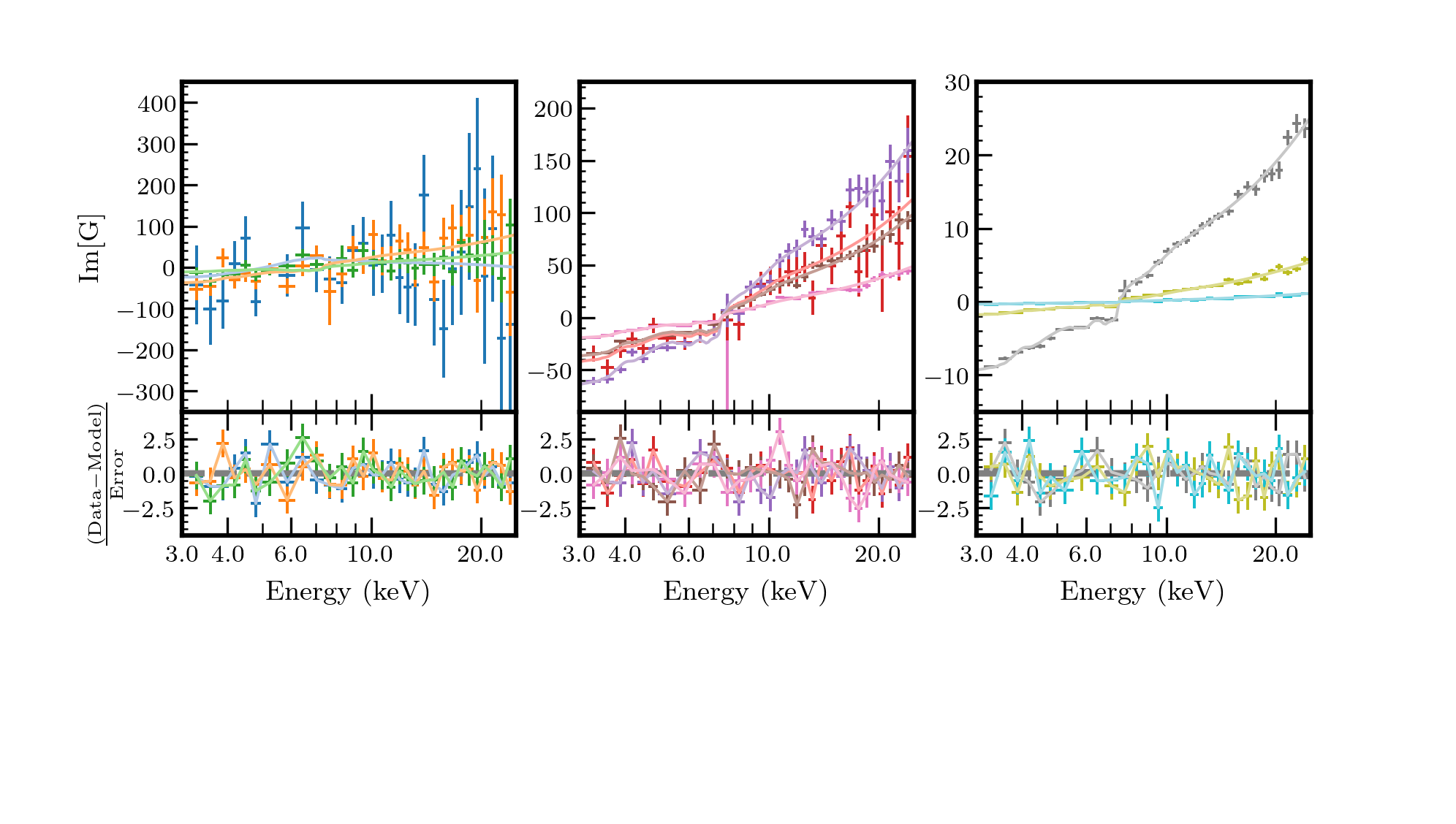}
    \vspace{-15pt}
    \caption[wide ref rtdist + extras cross spectrum]{Model C: Unfolded real (top panels) and imaginary (bottom panels) parts of the cross spectrum in all 10 frequency bands. The bottom axis of each panel show the contributions to $\chi^2$.}
    \label{fig:rtdist_cs}
\end{figure*}


\rev{Our initial fit features} structured residuals at 3-4 keV in the time-averaged and cross spectra (Fig. \ref{fig:comparison}, purple)\rev{, which}
we address
by including \rev{a} \texttt{diskbb} \rev{component} in the model for the time-averaged spectrum. This represents thermalised radiation from the disc. The cross spectra exhibit similar residuals, but it is somewhat unphysical to add a \texttt{diskbb} component to the cross spectrum, which would represent rapid variability of the emission\revnew{, correlated with no causal lag across}
the entire disc. It is much more likely that disc variability is dominated by radiation from the inner disc. We therefore add a blackbody component (\texttt{bbody}) to the cross spectral model. All cross spectrum blackbody temperatures were tied to one another, but were not tied to the time-average \texttt{diskbb} temperature. This means that the blackbody component represents variable blackbody emission from a single radius of the accretion disc. The \bbody temperature being equal to the \diskbb temperature corresponds to variability originating from $r_{\rm in}$, whereas the \bbody temperature being less than the \diskbb temperature corresponds to the variable radius being larger than $r_{\rm in}$. In reality, this approximates the variability coming from a narrow range of disc radii.

The \bbody normalization, $N_{\rm{BB}}$, is an independent free parameter for the real and imaginary parts of the cross spectrum of each frequency range. The normalization evolving with frequency is indicative of the power spectrum of disc variability. The normalization of the imaginary part of the cross spectrum in a given frequency range relates to the time lag between the disc variability and the reference band variability at that frequency. Specifically, the normalization of the imaginary part being larger than zero would indicate the disc variability lagging the variability in the reference band.

We also find evidence of a $\sim4$ keV calibration feature, similar to that previously reported by \cite{Garcia:2014} for the Crab and \cite{Connors:2020} for XTE J1550-564. \cite{Garcia:2014} developed the tool \texttt{pcacorr} to correct for such features, but unfortunately our data were taken too early in the RXTE mission for \texttt{pcacorr} to be applicable. We instead include a Gaussian absorption line with \texttt{gabs}. This feature could additionally be accounting for variable line-of-sight absorption, which is known to be present in the Cyg X-1 spectrum \citep{Lai:2022}. \rev{After extensive parameter exploration, we opt to freeze the centroid and width of the line to $4$ keV and $0.15$ keV respectively, but leave the line strength as a free parameter (tied across all spectra). We find that freezing the centroid and width hugely improves the stability of the fit without significantly influencing the other parameters. Fig \ref{fig:comparison} (orange) shows that the structured residuals are no longer present after the disc and Gaussian components are included.}

The model we use is: 
\begin{equation}
\label{eq:modelB}
\texttt{TBabs} \times \texttt{gabs} \times ( \texttt{diskbb}+\texttt{bbody}+\texttt{reltransDCp}).
\end{equation}
This is the model we fit to both the time averaged and cross spectrum simultaneously. The model is initialised such that the parts that affect the time-averaged spectrum are given by:
\begin{equation}
\label{eq:modelBTA}
\texttt{TBabs} \times \texttt{gabs} \times ( \texttt{diskbb}+\texttt{reltransDCp}),
\end{equation}
and the cross spectrum is affected by:
\begin{equation}
\label{eq:modelBCS}
\texttt{TBabs} \times \texttt{gabs} \times ( \texttt{bbody}+\texttt{reltransDCp}).
\end{equation}
The model parameters of \texttt{TBabs} and \texttt{gabs} are tied between the time averaged spectrum and all cross spectra. The \texttt{diskbb} normalization is a free parameter for the time averaged spectrum, and is frozen to zero for all cross spectra. Conversely, the \texttt{bbody} normalization is frozen to zero for the time-averaged spectrum and is free for the cross spectra. Within the \reltransdcp model, certain parameters are also tied between the time-averaged data and the cross spectra (such as lamppost height\footnote{See Table \ref{tab:params} for a full list of \reltransdcp parameters that are tied between time-averaged and cross spectra.}). However there are Fourier frequency dependent spectral pivoting parameters, $\phi_{\rm{AB}}$ and $\gamma$, that are used to model the cross spectrum in specific frequency bands. These parameters relate to the phase difference and amplitude ratio of spectral index and normalization variability respectively. For the time-averaged spectrum $\phi_{\rm{AB}} = \gamma = 0$; for the cross spectrum $\phi_{\rm{AB}}$ and $\gamma$ are free parameters that are tied between real and imaginary parts of the cross spectrum in a given frequency band. For example, we need 10 $\phi_{\rm{AB}}$ and $\gamma$ to model the real and imaginary parts of the cross spectrum (i.e. one set of $\phi_{\rm{AB}}$ and $\gamma$ are used to calculate the cross spectrum in a defined frequency range). We further discuss $\phi_{\rm{AB}}$ and $\gamma$ in Appendix \ref{sec:cross_spectrum}.

We obtain a good fit with
\rev{$\chi^2/{\rm dof} =  597/ 601$}; which we refer to as Model B. We list the best fitting parameters in Table \ref{tab:params}, and show our fit to the time averaged spectrum in Fig. \ref{fig:TA} (orange points). Fig. \ref{fig:AIDCP} shows the fits to all cross spectra. We show real (top) and imaginary (bottom) parts of the cross spectrum in each of the 10 frequency bands (as labelled). The data are unfolded around the best fitting model.
In Fig. \ref{fig:dcp_lag_amp}, we convert the real and imaginary parts of the cross spectra into time lag (top) and absolute variability amplitude (bottom). The data were converted from the unfolded real and imaginary parts of the cross spectra. We do not plot the lag data for the lowest frequency ranges due to the \revnew{errors} being very large. Although the real and imaginary parts of the cross spectrum were used in the fit (for statistical reasons detailed in \citealt{Mastroserio:2018}), the lag and amplitude are easier to interpret physically. We see a dip at the iron line of the model lag spectra, which is due to the reverberation lag being smaller than the continuum lag \citep{Mastroserio:2018}. We also see the reflection features in the amplitude spectrum becoming less prominent with increasing frequency, which is due to the finite size of the reflector washing out fast variability in the reflected signal via path length differences \citep{Revnivtsev:1999,Gilfanov:2000}.
 
\subsection{Model C: \rev{\rtdist}}
\label{sec:modelC}
\begin{figure*}
    \centering
    \includegraphics[trim={0cm 0.cm 0.cm 0.cm},clip,width=\linewidth]{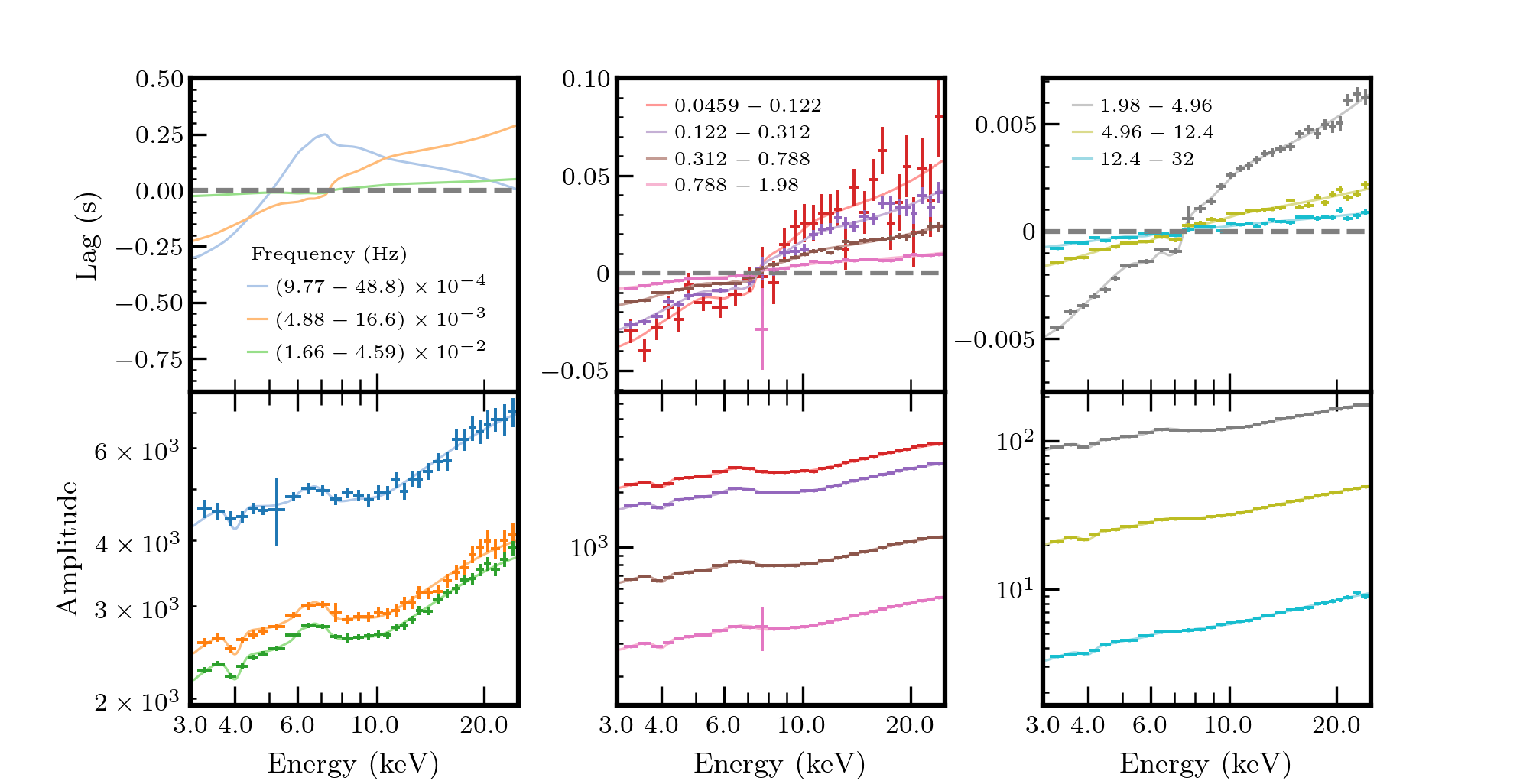}
    \caption{Model C: Lag-energy spectrum (top) and variability amplitude (bottom panels) for different Fourier frequency bands. The points are calculated from the unfolded cross spectra and lines are calculated using parameters of model C. We only show the model lines for the smallest frequency bands, the data in these frequency bands are not shown due to their large \revnew{errors}. The units of amplitude are keV$\,$cm$^{-2}\,$s$^{-1}$.}
    \label{fig:rtdist_la}
\end{figure*}
We now replace \reltransdcp\ used in model B with \rtdist. 
The headline difference between these two models is that \rtdist\ takes distance $D$ as an input in place of the peak ionization parameter $\log\xi$, which is instead self-consistently calculated \citep{Ingram:2022}. There are some other more subtle differences between \reltransdcp\ and \rtdist. Most importantly, the \texttt{boost} parameter is treated differently. 
In \texttt{reltransDCp}, the relative normalization of the direct and reflected components is set to the value calculated self-consistently for a stationary, isotropically emitting lamppost corona multiplied by \texttt{boost} \citep{Ingram:2019a}. 
In this way, the amount of reflection in the spectrum is artificially increased from the calculated value if $\texttt{boost}>1$. 
This is designed to account for the isotropic lamppost assumption surely being an over simplification. 
In \rtdist\, the \texttt{boost} parameter instead adjusts the angular emissivity of the lamppost corona. 
Setting \texttt{boost}$>1$ now beams radiation towards the black hole, which increases the amount of reflection in the output spectrum \textit{and} leads to reflection being more centrally concentrated on the disc (i.e. the radial emissivity profile becomes steeper).
This is in principle a more self-consistent treatment than that employed in the earlier \reltrans\ models. 
The \rtdist\ model also includes the parameters $b_1$ and $b_2$, which further distort the angular emissivity profile of the corona away from isotropic if they are non-zero. 
For $\texttt{boost}=1$ and $b_1=b_2=z/r=0$, the two models are able to output identical results, as long as the user finds the value of $D$ in \rtdist\ that corresponds to the peak ionisation parameter used for \reltransdcp, otherwise the two models otherwise diverge.
We see from Table \ref{tab:params} that the best fitting Model B $\texttt{boost}$ parameter is $\texttt{boost} \approx 0.29$, and so we already know that we will not be able to exactly reproduce the Model B fit with \rtdist\ in place of \reltransdcp.

We use the same procedure described in the previous subsection to implement the model
\begin{equation}
    \texttt{TBabs} \times \texttt{gabs} \times ( \texttt{diskbb}+\texttt{bbody}+\texttt{rtdist}),
\end{equation}
and we refer to our best fit using this as Model C. As was the case with Model B, the calculated time-averaged spectrum and cross spectra are affected by parameter changes in the components outlined in equations \ref{eq:modelBTA} \& \ref{eq:modelBCS} respectively, albeit \reltransdcp is now replaced with \rtdist. Our best fit is \rev{$\chi^2 = 645/599$}; we show our fit to the time average spectrum in Fig. \ref{fig:TA} (green); the cross spectra in Fig. \ref{fig:rtdist_cs}; and list our best-fitting parameters in Table \ref{tab:params}. In Fig. \ref{fig:rtdist_la}, we again convert the real and imaginary parts of the cross spectra to time lag and variability amplitude. We again see the dip at the iron line of the lag models and the weakening of reflection features with increasing frequency in the amplitude spectrum.

\begin{table*}
\begin{tabular}{llll} \toprule
Model & Model A & Model B & Model C \\
\midrule
Absorption Model & \texttt{TBabs}& \texttt{TBabs} & \texttt{TBabs} \\ 
\midrule
$N_H$ / $10^{22}$ & $1.30^{+0.08}_{-0.13}$& $3.0 ^{+ 0.5 }_{ -0.4 }$ & $ 2.6 ^{+ 0.1}_{ -0.1}$ \\
\toprule
Absorption Line Model & & \texttt{gabs}& \texttt{gabs} \\ \midrule
E / keV & --- & $4.00 $ & $4.00 $ \\
Width / keV & --- & $0.15$ & $0.15$ \\
Strength & --- & $0.032 ^{+ 0.005 }_{ -0.004 }$ & $0.038 ^{+ 0.004 }_{ -0.004}$ \\
\toprule
Time-average Blackbody Model &  & \texttt{diskbb} & \texttt{diskbb}\\ \midrule
$kT_{in}$ / keV & --- & $0.71 ^{+ 0.03 }_{ -0.03}$ & $0.92 ^{+ 0.03}_{ -0.03}$ \\
normalization & --- & $700 ^{+ 220 }_{ -160 }$ & $160^{+ 30}_{ -20}$ \\
\toprule
Cross spectrum Blackbody Model &  & \texttt{bbody} & \texttt{bbody}\\ \midrule
$kT$ / keV & --- & $0.640^{+0.02}_{-0.004}$ & $0.69^{+0.02}_{-0.01}$ \\
\toprule
Reflection Model & \reltransdcp & \reltransdcp & \rtdist \\ \midrule
$h$ / $R_g$ & $10.5^{+0.5}_{-0.8}$& $11.4 ^{+ 1.1}_{ -1.0 }$ & $9.9 ^{+ 0.7 }_{ -0.6}$ \\
$i$ / $^{\circ}$ & $37^{+1}_{-1}$& $ 42 ^{+ 2 }_{ -2}$ & $43 ^{+ 2 }_{ -2 }$ \\
r$_{\rm in}$ / $R_g$ & $4.7^{+0.7}_{-0.9}$& $5.7^{+0.2}_{-0.2}$ & $18.8 ^{+ 1.4}_{ -1.3}$ \\
$\Gamma$ & $1.737^{+0.007}_{-0.004}$& $1.705 ^{+ 0.007 }_{ -0.006}$ & $1.706^{+0.006}_{-0.006}$ \\
$D_0$ / kpc & --- & --- & $3.9^{+0.5}_{-0.5}$ \\
log $\xi$ & $3.40^{+0.03}_{-0.01}$& $3.40 ^{+ 0.03 }_{ -0.03 }$ & --- \\
$A_{Fe}$ & $1.8^{+0.2}_{-0.1}$& $2.6 ^{+ 0.3 }_{ -0.2 }$ & $1.9^{+0.2}_{-0.2}$ \\
log $n_e$ & $15.0$& $19.0 ^{+ 0.3 }_{ -0.3 }$ & $19.1^{+0.2}_{-0.2}$ \\
$kT_e$ / keV & $140^{+90}_{-50}$& $360 ^{+ 100 }_{ -100}$ & $380^{+70}_{-70}$ \\
boost & $0.37^{+0.01}_{-0.01}$& $0.29 ^{+ 0.02 }_{ -0.02 }$ & $0.7^{+0.1}_{-0.1}$ \\
$b_1$ & --- & --- & $1.5^{+0.1}_{-0.1}$\\
$b_2$ & --- & --- & $-3.7^{+0.5}_{-0.3}$\\
$M$ / M$_{\odot}$ & $25^{+7}_{-7}$& $15 ^{+ 4 }_{ -4 }$ & $9.5^{+2}_{-2}$ \\
normalization & $0.092^{+0.007}_{-0.004}$ & $0.100 ^{+ 0.004}_{ -0.004 }$ & $0.108^{+0.004}_{-0.004}$\\
\toprule
$\chi^2$ & $492/563$ & $597/601$ & $645/599$\\
\bottomrule
\end{tabular}
\caption{Best fitting parameters obtained from $\chi^2$ minimization from to the cross spectrum in 10 frequency ranges ($0.98$ mHz - $32$ Hz) and the time-average spectrum. In both models the spin was fixed to 0.998. In \reltransdcp distance is not a parameter hence is not included. In \rtdist, log$\xi$ is internally calculated hence is not a free parameter. Errors are all $90$ per cent confidence. We do not list the empirical parameters  for each frequency band, i.e. the \texttt{bbody} normalizations ($N_{\rm BB}$) for real and imaginary parts, and the \reltrans continuum lag parameters $\phi_{\rm{AB}}$ and $\gamma$. We instead summarise the empirical parameters in Appendix \ref{sec:cross_spectrum}.}
\label{tab:params}
\end{table*}

\section{Results}
\label{sec:results}

Table \ref{tab:params} lists the best-fitting parameter values of all three models alongside the corresponding fit statistic. For Model A we obtain parameter uncertainties using the \textsc{xspec} \texttt{error} command, whereas for Models B and C we estimate errors from Monte Carlo Markov Chain (MCMC) simulations. We initialise the MCMC simulation from the best fitting Models B and C found via $\chi^2$ minimization. \rev{The Model B and C chains both have total length 307,200 for 256 walkers (i.e. 1,200 steps per walker) after an initial burn-in phase of 19,968 and 4,285,052 for Models B and C respectively. The Model C burn-in is required to be much longer because its parameter space is much more complex.}
For each parameter in both chains we find the Geweke convergence measure to be within the range -0.2 to 0.2, indicating convergence. 

\rev{We adopt Model B as our best fitting model, since it yields a significantly better fit than Model C (Model A employs over-estimated errors, and is intended purely as a consistency check). Model B has two fewer free parameters than Model C, yet has a lower $\chi^2$ by $\Delta\chi^2 = 48$. We therefore primarily consider Model B in this section, with some comparisons to Model C to explore the dependence of inferred parameters on model assumptions.}


\subsection{Mass and distance}
\label{sec:MandD}


\begin{figure*}
    \centering
    \includegraphics[width=0.8\linewidth]{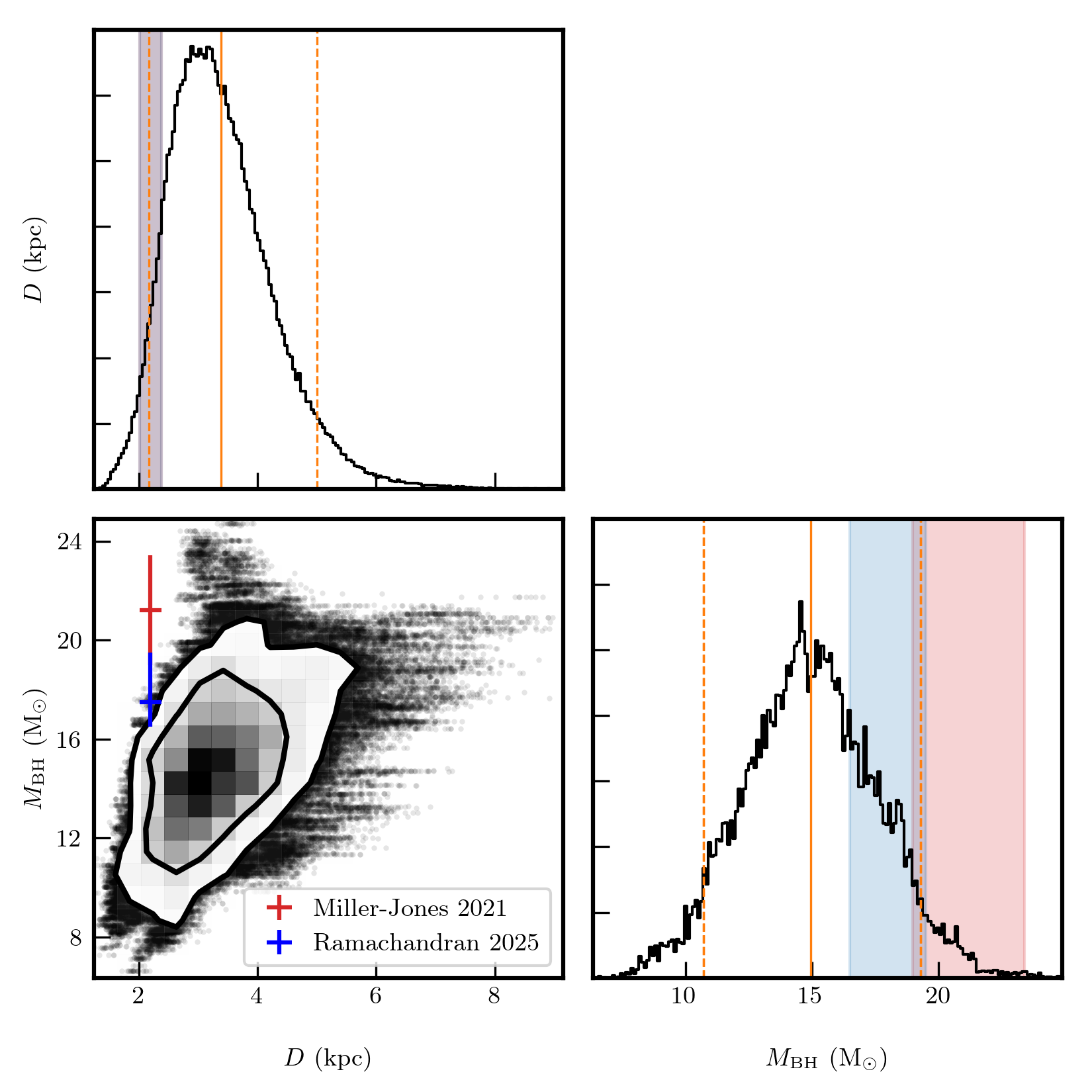}
    \vspace{-10pt}
    \caption[Steppar analysis]{\rev{2D distance and mass contour for Model B (lower-left), where the contours are the 2D 68 and 90 per cent confidence intervals, \revnew{and the pixels show the kernel density estimate of the MCMC samples. The points outside the 90 percent interval show a streak-like pattern resulting from our correction for the absolute flux uncertainty (see text in Section \ref{sec:MandD}).} We also show the 1D distance and mass distributions (upper-left and lower-right respectively) alongside the measured values obtained in \cite{MillerJones:2021} and \cite{Ramachandran:2025}, indicated by the red and blue shaded regions respectively. For both histograms we show the mean position and 1D 90 percent confidence interval with orange solid and dashed lines respectively.}}
    \label{fig:DM_error}
    
\end{figure*}

\revnew{Black hole mass is a parameter in all models considered. However only Model C, which uses \rtdist, takes distance as an input parameter, though distance can also be inferred from models implementing \reltransdcp}. We use a post-processing code to convert the peak ionization parameter of \reltransdcp to distance. The code uses Equation (10) of \cite{Ingram:2022}, which relates the ionization parameter $\xi(r)$ to the distance via the radial emissivity profile, $\epsilon(r)$. For \rtdist, $\epsilon(r)$ is defined by Equation (6) of \cite{Ingram:2022}. In our post processing code, we instead use the definition of $\epsilon(r)$ from Equation (20) of \cite{Ingram:2019a}, multiplied by \texttt{boost}. This the emissivity profile that is employed in \reltransdcp, which assumes a stationary, isotropic lamppost before artificially adjusting the reflected flux by a factor \texttt{boost}. The post processing code, which \rev{is publicly available in version 2.3 onwards}
of the \reltrans package, reads in an MCMC table output from \reltransdcp fits and adds a column consisting of a distance value for each step in the chain.

For both Models B and C, we apply further post processing to the distance values to account for absolute flux uncertainty 
\citep[following][]{Nathan:2024}. For each step in the chain, we estimate the true distance $D$ as
\begin{equation}
    D = \frac{D_0}{\sqrt{x (1+\eta)}},
\end{equation}
where $D_0$ is the `raw' distance measurement output by the model, $x$ is a constant that accounts for the known systematic flux offset of the instrument, and $\eta$ is a Gaussian random variable with zero mean and a standard deviation of $\sigma$ that accounts for calibration uncertainty. We use $x=9.5/10.7$, since \cite{Garcia:2014} measure the normalization of the power-law spectrum of the Crab nebula to be $\approx 10.7$ for the earliest epochs of the RXTE mission (which are most relevant to our dataset), whereas the accepted value is $\approx 9.5$ \citep{Toor1974}. We set $\sigma=0.1$ to conservatively employ a $10$ per cent absolute flux uncertainty for RXTE, following e.g. \cite{Steiner:2012}. Note that flux $\propto 1/D^2$, thus the inferred distance goes as one over the square root of flux. This means that a $10$ per cent flux uncertainty translates to a $5$ per cent distance uncertainty.

Fig. \ref{fig:DM_error} shows the resulting joint mass-distance posterior for Model B\rev{, alongside the dynamical and parallax measurements of \cite{MillerJones:2021} and \cite{Ramachandran:2025}. We see that our best fitting model agrees with the \cite{Ramachandran:2025} values within the $90$ per cent 2D statistical confidence interval, even when we do not take into account the effect on the dynamical mass measurement of uncertainty on the inclination angle. Our best fitting mass and distance (with $90$ per cent uncertainties) are $M=15\pm 4~M_\odot$ and $D=3.4_{-1.2}^{+1.6}$ kpc. Model C yields a mass of $M=9.5\pm 2~M_\odot$ and a distance of $D=3.9\pm 0.5$ kpc, which are each consistent with the corresponding Model B values within 90\% confidence  (albeit for a far larger $\chi^2$).}



\subsection{Disc variability}
Models B and C include a \texttt{diskbb} component in the time averaged spectrum and a \texttt{bbody} component in the cross spectra. For the best fit of both models, the \texttt{bbody} temperature is slightly lower than the peak \texttt{diskbb} temperature. This corresponds physically to a disc that is accreting stably through most of its extent but exhibits variability at some narrow range of radii in its inner regions, as suggested by e.g. \cite{Wilkinson:2009,Rapisarda:2016}. \revnew{As the flux of a blackbody component fluctuates, so too does its temperature via the Stefan-Boltzmann law. The resulting non-linear variability of the blackbody function can be linearised via a first-order Taylor expansion to result in a constant component that peaks at $E=2.82~kT$ (i.e. the Planck function), plus a variable component that peaks at $E=3.82~kT$; i.e. a factor 1.35 higher \citep{vanParadijs:,Uttley:2025}. In our best fitting model we measure $kT\approx 0.64$ keV for the variable component, which corresponds to an equilibrium temperature of $kT_0 \approx 0.64/1.35=0.47$ keV.} Since the disc temperature goes as $T_0/T_{\rm in} = (r/r_{\rm in})^{-3/4}$ in the \texttt{diskbb} model, we can estimate that the outer radius of the variable region in the disc is $(T_{\rm in}/T_0)^{4/3}~r_{\rm in}\sim \revnew{1.7}~r_{\rm in}$ for both models.

\begin{figure}
    \centering
    \includegraphics[width=\linewidth]{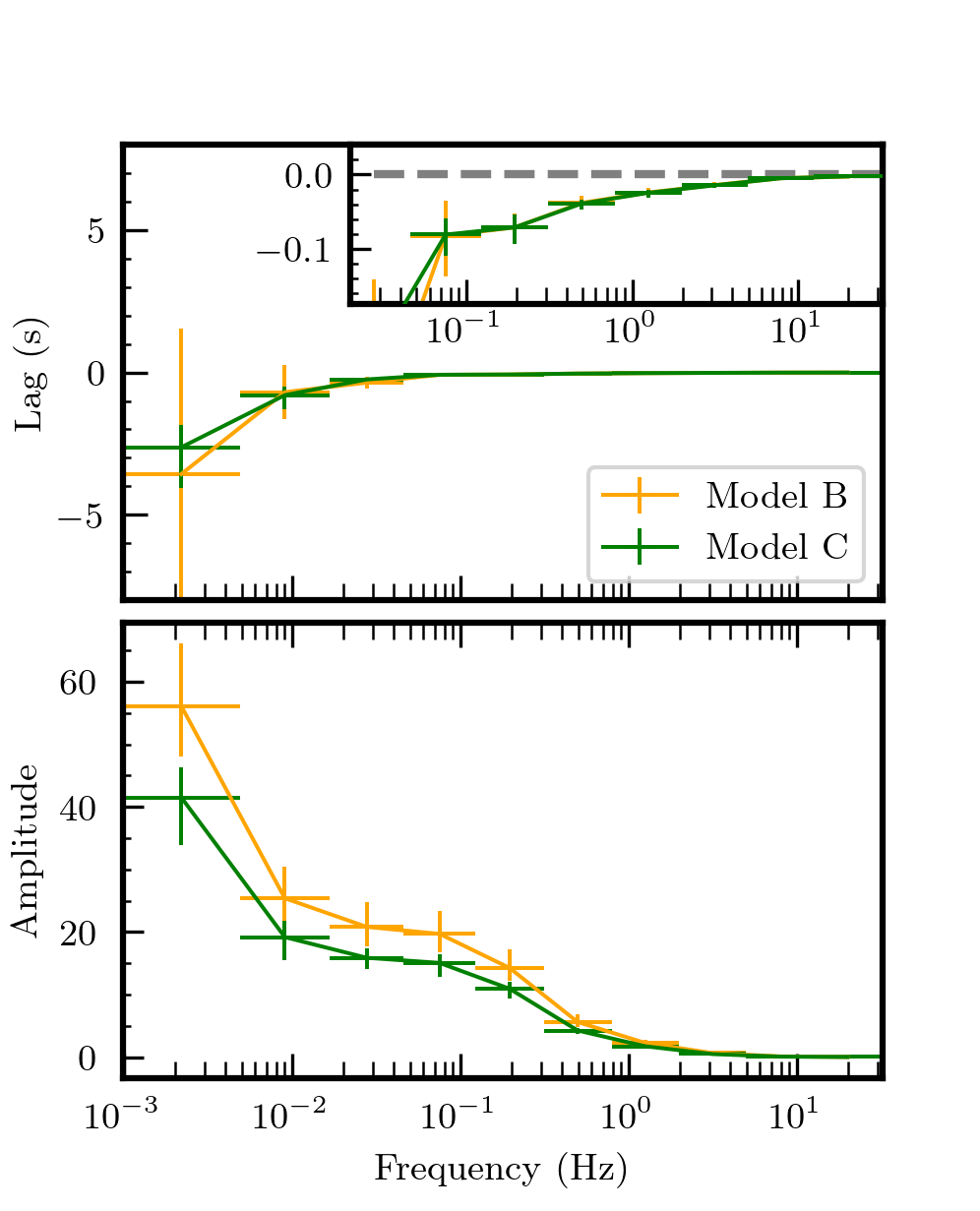}
    \vspace{-25pt}
    \caption{Lag-frequency spectrum (\textit{top}) and variability amplitude (\textit{bottom}) associated with the \texttt{bbody} obtained from the models B \& C with our best fit parameters. The inset axis in the top panel is the zoomed-in region of the higher frequency lags. We measure negative lags, indicating that variability in the blackbody component leads that in the reference band. \revnew{Errors} shown are the 90 per cent confidence interval. The units of amplitude are the same as the units of the blackbody normalisation parameter.}
    \label{fig:thermal_lags}
\end{figure}

\begin{figure}
    \centering
    \includegraphics[width=\linewidth]{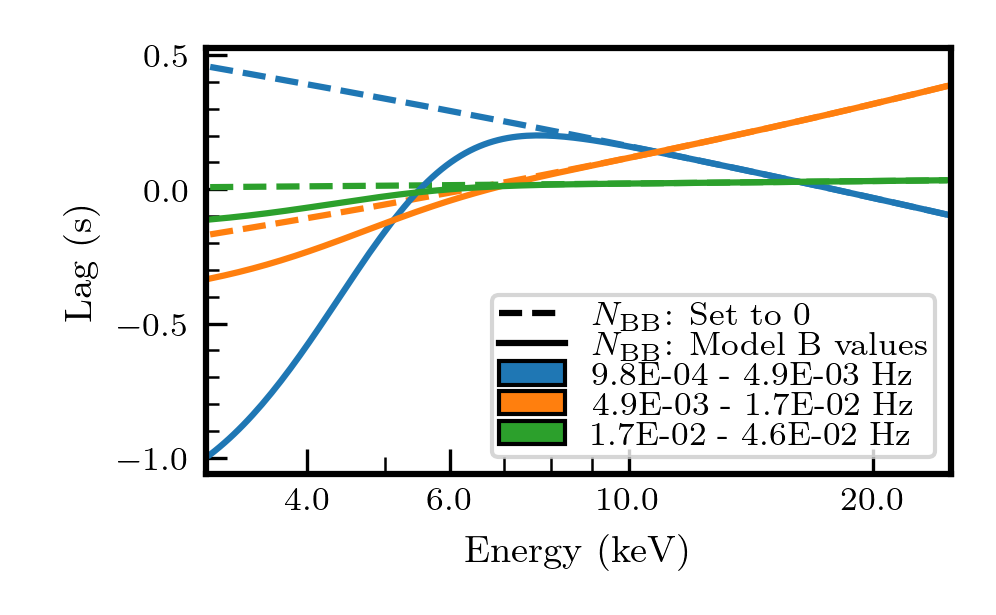}
    \vspace{-20pt}
    \caption[wide ref dcp + extras cross spectrum]{Collection of the model predicted lag-energy spectra for Model B in the lowest frequency bands, with reflection switched off (\texttt{boost}=0). Since reflection is switched off, there are no reverberation lags. The dashed lines are when disc variability is also switched off, leaving only the log-linear continuum lags. The solid lines include disc variability, which we see contributes a negative lag at lower energies.}
    \label{fig:continuum_lags}
\end{figure}

In Fig. \ref{fig:thermal_lags}, we show how the best fitting \texttt{bbody} normalization parameters depend on Fourier frequency. The parameters we fit for are the normalizations of the real and imaginary components, $N_{\rm {BB}, \rm{real}}$ and $N_{\rm {BB}, \rm{im}}$. However, it is more instructive to plot in terms of amplitude and phase. The bottom panel shows variability amplitude, $\sqrt{N_{\rm {BB}, \rm{real}}^2+N_{\rm {BB}, \rm{im}}^2}$. We see that this decreases with Fourier frequency, which is expected since high frequency variability is expected to be damped by viscous processes \citep[e.g.][]{Lyubarskii:1997, Frank:2002}. The top panel shows time lag, $\arctan(N_{\rm {BB}, \rm{im}}/N_{\rm {BB}, \rm{real}})/(2\pi\nu)$. Due to the sign convention, this shows that disc variability leads variability in the reference band. This is again expected physically for disc variability driven by accretion rate fluctuations, since those fluctuations will propagate inwards on a viscous timescale before modulating the coronal flux \citep{Lyubarskii:1997,Uttley:2025}.
Fig. \ref{fig:continuum_lags} demonstrates the influence of disc variability on the overall time lags in the model. Since disc variability is only prominent at low frequencies, we only plot for three low frequency ranges. The dashed lines show the lags resulting from spectral pivoting alone; i.e. the reflection and disc components have both been switched off. We see only the log-linear lags that result from spectral pivoting \citep{Kotov:2001,Mastroserio:2018}. The solid line is still without reflection, but now includes disc variability as well as spectral pivoting. We see that the disc contributes a negative lag at soft X-rays, due to disc variability leading coronal variability. Note that here we plot only for Model B, since for \reltransdcp it is easy to switch off the reflection component, but very similar properties are also seen for Model C. Details of the \reltrans spectral pivoting parameters $\phi_{AB}$, $\gamma$; and \bbody parameters $N_{\rm {BB}, \rm{real}}$, $N_{\rm {BB}, \rm{im}}$ is given in Appendix \ref{sec:cross_spectrum}.   

\subsection{Other parameters}

We see from Table \ref{tab:params} that several parameters are consistent across all three models, including the source height $h \sim 10~R_g$. The inclination varies moderately across the three models from $i \approx 37^\circ$ to $i \approx 43^\circ$, all of which are larger than the binary inclination \revnew{of $i \approx 27.5^\circ$} \rev{inferred by \cite{MillerJones:2021}}. \revnew{We trialled the Model B configuration while freezing $i=27.5^\circ$, which resulted in a significantly larger $\chi^2 = 666/602$. We note, however that in this new fit the disk density was pegged to the \texttt{xillver} upper limit of $n_e=10^{20}$ cm$^{-3}$, implying that a better fit could potentially be found for $i=27.5^\circ$ if it were possible to push \texttt{xillver} to higher densities.}


One striking difference between Models B and C is the disc inner radius, which is $r_{\rm in} \approx 6~R_g$ for Model B and \rev{$r_{\rm in} \approx 19~R_g$} for Model C. This emphasises that the measured inner radius can be very dependent on model assumptions, as has been discussed many times previously \citep[e.g.][]{Shreeram:2020, Basak:2017, Zdziarski:2021,Zdziarski:2022}. \rev{The difference in $r_{\rm in}$ in this case likely results from Models B and C having very different radial emissivity profiles to one another, due to the different implementation of the boost parameter and the use of the anisotropy parameters $b_1$ and $b_2$ in Model C.} We note, however, that Model B has a much lower $\chi^2$ than Model C.

\rev{High disc density is returned by both models for which it is a free parameter, as is expected for a stellar-mass black hole \citep{Garcia:2016,Tomsick:2018}. Mildly super-solar values of iron abundance $A_{\rm Fe}$ are returned for all models, as is commonly found \citep{Garcia:2018}.}
Finally, the electron temperature $kT_e$ is a lot higher for Model \rev{B \&} C compared with \rev{Model A}. \revnew{The PCA data we consider here} only extend to $25$ keV, so we are not directly sensitive to $kT_e$, but we \textit{are} indirectly sensitive through its influence on the shape of the restframe reflection spectrum \citep{Garcia:2015}. \revnew{We investigated the electron temperature further by including High Energy Timing Experiment (HEXTE) data into our Model B fit. We considered the $50-200$ keV HEXTE spectrum 
taken during the first of the five observations for an exploratory fit. We included a cross-calibration constant into the model that we froze to unity for the PCA and left as a free parameter for HEXTE (following, e.g. \citealt{Barrio:2003}). 
Fitting this model yielded $\chi^2 = 624.51/629$ and $kT_e = 360 \pm 90$ keV, consistent with the Model B value given in Table \ref{tab:params}.}


\section{Discussion}
\label{sec:discussion}

We have conducted an X-ray reverberation mapping analysis on RXTE data of Cyg X-1 in the hard state using models from the \reltrans package. This has enabled us to make the first estimate of distance to a BHXB through X-ray reverberation mapping.

\subsection{Disc variability}

This work builds upon earlier work on the same dataset by M19. One motivation to revisit the dataset was to address the problem presented by unphysically large \revnew{errors} that were implemented during the M19 analysis. Using the error formulae for the cross spectra from \cite{Ingram:2019b}, coupled with the extension of the low band pass considered for our fit from 4 to 3 keV, reveal the need for disc variability. Our results are consistent with the disc accreting stably outside of $r \sim \revnew{1.7}~r_{\rm in}$, with accretion rate variability becoming prominent inside of that radius. We find that the disc variability leads variations in the reference band, which is dominated by the corona. This is consistent with mass accretion rate variations propagating inwards from the disc before eventually reaching the corona after a propagation time \citep{Wilkinson:2009,Rapisarda:2016}. Fig. \ref{fig:thermal_lags} (top) shows that the observed lag is $\sim 1$ s at the lowest frequencies, and gets shorter with increasing frequency. This reduction of the lag with increasing frequency occurs in the propagating fluctuations model because the characteristic variability frequency generated in the disc -- the \textit{viscous frequency} -- increases with decreasing radius \citep[e.g.][]{Lyubarskii:1997,Churazov:2001,Frank:2002,Mushtukov:2018}. Since the highest frequency variability can only be generated in the inner disc, it cannot have propagated far. On the other hand, low frequency variability can be generated further out in the disc and thus the observed low frequency lags include contributions from variability that has propagated a long distance \citep{Arevalo:2006,Ingram:2013}. Thus, if the observed disc variability is indeed driven by accretion rate fluctuations, the true propagation lag from the edge of the variable region ($r \sim \revnew{1.7} ~r_{\rm in}$ in this case) to the corona is longer than the observed lag of $\sim 1$ s.

Fig. \ref{fig:thermal_lags} (bottom) shows that the amplitude of disc variability is roughly constant at lower frequencies (except for the lowest frequency band, which shows an excess), and starts to drop off above $\nu \sim 0.1$ Hz. This drop off is again expected in the propagation fluctuations model, and can be interpreted as a lower limit for the local viscous frequency in the variable region of the disc. The viscous frequency at $r_{\rm in}$ sets the fastest characteristic variability timescale in the disc mass accretion rate. The observed radiation is the sum of that emitted from different regions of the disc, and interference between signals from different regions washes out variability at the highest frequencies, leaving the observed break at a frequency lower than the peak viscous frequency \citep{Ingram:2013}. \revnew{Our treatment is a rather crude approximation of this process, consisting of a constant disc component plus a single blackbody function that represents the variable disc region. From the best fitting \texttt{diskbb} and \texttt{bbody} temperatures, we estimate that the fractional rms of this single variable region needs to be $\sim 45\%$ if the region extends from $r_{\rm in}$ to $1.7~r_{\rm in}$ (see Appendix \ref{sec:discrms} for the derivation).} In future it may be possible to quantitatively apply a detailed propagating fluctuations model to the disc variability observed here \citep[following e.g. ][]{Rapisarda:2017}.

Disc variability has also been suggested to drive the fluctuations in the power law index of the corona that the \reltrans models input in an \textit{ad hoc} manner to reproduce the continuum lags. \cite{Uttley:2025} showed that an over-density in the disc first leads to cooling of the corona via an increase in seed photons, then after a propagation time heats the corona (also see \citealt{Karpouzas:2020}). The instantaneous power law index can be approximated as $\Gamma(t) \approx \Gamma_0 [ L_s(t) / L_h(t) ]^{1/6}$ \citep{Beloborodov:2001}, where $\Gamma_0$ is the time-averaged photon index, $L_s(t) \propto \dot{M}(\revnew{1.7}~ r_{\rm in},t)$ is the seed photon luminosity and $L_h(t)  \propto \dot{M}(\revnew{1.7}~ r_{\rm in},t-\Delta t)$ is the coronal heating luminosity. Whereas here we simply parameterise $\Gamma(t)$ fluctuations and their delay with respect to luminosity variations with the parameters $\gamma$ and $\phi_{AB}$ (see Fig. \ref{fig:empirical}), in future we may be able to self-consistently calculate $\Gamma(t)$ variations from disc variability in the framework of the \cite{Uttley:2025} model.

\subsection{Geometrical parameters}

We fit three models, referred to as Models A, B and C (see Table \ref{tab:params}). Model A is purely intended to verify consistency with the earlier work of M19, which it does successfully. Models B and C are the fits worth further discussion. These two models differ only in their treatment of the reflection fraction, which is controlled in both cases by the \texttt{boost} parameter. 
The meaning of this parameter is quite different for the two models. Model B uses \reltransdcp \citep{Mastroserio:2021}, which calculates the reflection fraction self-consistently for a stationary, isotropic lamppost source before artificially adjusting the normalization of the reflection spectrum by a factor \texttt{boost}. This adjustment is intended to account for the over-simplified nature of the lamppost model. \rev{Essentially, \reltransdcp uses the lamppost geometry as a mathematical approximation, and adjusts the normalisation to account for this being an oversimplification of the true geometry.}

For Model C, which uses \rtdist \citep{Ingram:2022}, the \texttt{boost} parameter adjusts the angular emissivity of the corona (i.e. boosts emission towards or away from the black hole). Thus \texttt{boost} $<1$ reduces the reflection fraction from the simple isotropic case for both models, but for Model B the shape of the reflection spectrum is unchanged, whereas for Model C the iron line is slightly narrower due to the coronal emission being boosted away from the black hole, thus resulting in less centrally peaked illumination of the disc. \rev{Thus, \rtdist is a self-consistent implementation of the lamppost model, intended as a realistic approximation of a very compact corona. It is therefore interesting that Model B returns a far better fit than Model C, essentially disfavouring a geometry whereby the corona is very compact and located on the black hole spin axis.}

Indeed, recent IXPE observations of the 2--8 keV polarization aligning with the jet \citep{Henric:2022} indicate that the corona is extended in the disc plane, which is formally incompatible with the lamppost model we employ here. \rev{It is not trivial to predict how using an extended corona geometry in place of a lamppost would change our results. If the radial extent of the corona is small compared with the disc inner radius,} it may predict light crossing delays comparable to those predicted by the lamppost model. \rev{Alternatively, if the corona is highly extended and sandwiches the disc,} the light crossing delays will be drastically different to those predicted by the lamppost model.


\rev{Models B and C}
yield quite different measurements of disc inner radius: $r_{\rm in}\approx 6~R_g$ for Model B and
\rev{$r_{\rm in}\approx 19~R_g$}
for Model C. In the truncated disc model, $r_{\rm in}$ is expected to be reasonably large in the hard state (consistent with our Model C); as is seen, for example, by \emph{Hard X-ray Modulation Telescope} observations of Cyg X-1 \citep{Feng:2022}; where they find the truncation radius decreases as the source softens. On the contrary, several reflection modelling studies find that the disc already reaches the ISCO in the bright hard state \citep{Garcia:2015,Liu:2023}. Model C is certainly consistent with the truncated disc picture, \rev{but it yields a significantly worse fit than Model B.}
The Model B value, although smaller, is still outside of the ISCO for a rapidly spinning black hole, which Cyg X-1 has been suggested to be (\citealt{Zhao:2021}, but see \citealt{Belczynski:2024} \rev{and \citealt{Zdziarski:2024}} for a contrasting view). However, our results here cannot contribute to the debate of whether the disc reaches the ISCO in the bright hard state, since Cyg X-1 is thought to always be on the lower branch of the \rev{hardness intensity diagram} (luminosity $\sim 2$ per cent Eddington). The sensitivity of the inferred disc inner radius to assumptions that we experience in our fits is representative of many other studies in the literature (e.g. compare \citealt{Buisson:2019} with \citealt{Zdziarski:2021}; or \citealt{Parker:2015} with \citealt{Basak:2017}).

Our fits yield an inclination angle slightly higher than the known binary system inclination: $i \approx \rev{42}^\circ$ and $i \approx \rev{43}^\circ$ for Models B and C compared with binary inclination of $i \approx 27.5^\circ$ \revnew{inferred from optical observations} \citep{MillerJones:2021}. \revnew{In fact, $i=27.5^\circ$ was strongly disfavoured by our fits}. This is consistent with results of previous time-averaged spectral studies, both in the soft state ($i>38^{\circ}$: \citealt{Tomsick:2014,Walton:2016}) and in the hard state ($i>42^{\circ}$: \citealt{Parker:2015}). Although the inclination angle inferred from reflection modelling is subject to systematic uncertainty, it is interesting to consider the case for Cyg X-1 having a warped disc, such that we see the inner disc and corona from a higher inclination than the outer disc, which presumably aligns in the binary plane. Indeed, such a misalignment is required for simple Comptonising slab corona models to explain the unexpectedly high 2--8 keV polarisation degree of $4$ per cent
\citep{Henric:2022}. \rev{One}
alternative explanation that allows for an aligned system invokes a $\sim 0.4 c$ bulk outflow velocity of the electrons in the corona \citep{Poutanen:2023}. In principle it is possible to obtain an outflow velocity from the \texttt{boost} parameter for Model C, however it is not so straightforward. What we can say is that, \texttt{boost} $<1$ is consistent with an outflow, but we cannot be more quantitative because the influence of an outflow on the emissivity profile will be sensitive to the unknown true coronal geometry.


\subsection{Mass and distance}


\rev{
For our best-fitting model, Model B (\reltransdcp), we measure a black hole mass of $M = 15\pm 4~M_\odot$ and a distance of $D = 3.4^{+1.6}_{-1.2}$ kpc, consistent with previous measurements \citep{Ramachandran:2025}. Model C (\rtdist) returns a much worse fit, but values $M$ and $D$ that both agree with our best fit values within 90 per cent confidence bounds.}

\rev{It is encouraging that our model returns such reasonable values for these parameters as}
there are many approximations in the models that could have dramatically affected accuracy. For example, the illuminating spectrum is calculated with \texttt{nthcomp} \citep{Zdziarski:1996}, with the seed photon temperature that determines the low energy cut off hardwired to $0.01$ keV\footnote{Listed as a hidden parameter for CP models on the \textsc{relxill} software
\href{https://www.sternwarte.uni-erlangen.de/~dauser/research/relxill/}{homepage} (under model parameters)
}. This value is unrealistically low (if the seed photons for Comptonisation are provided mainly by the disc), and so the disc ionization state is calculated within the model assuming the wrong illuminating spectrum. In future, including the seed photon temperature as a free parameter would require new, extended \texttt{xillver} grids, but would mean the model calculation will be a better representation of the reflection process.



\rev{\subsection{\rtdist bug}}

\rev{During our analysis, we discovered and fixed a bug in the \rtdist source code. We found that the flux was multiplied by an erroneous factor of $4\pi$ within the code, meaning that the distance being output was biased by a factor of $\sqrt{4\pi}$ (i.e. to get the true distance, one must multiply the distance output from an old version of \rtdist by $\sqrt{4\pi}$). The distances we present in this paper have been corrected for this issue by scaling the output $D_0$ parameter by $\sqrt{4\pi}$. The mistake is no longer present from version 2.3 onwards of the \reltrans package.

The bug only materially affects the conclusions of one published paper \citep{Nathan:2024}. It is also present in the analysis of \cite{Ingram:2022}, but does not affect the conclusions because the same model was used both to create synthetic data and to fit back to that fake dataset. The bug is also in the neural network emulator \texttt{rtfast} \citep{Ricketts:2024}, which is trained on the old version of \rtdist. Users of the affected \texttt{rtfast} version simply need to multiply the distance output by the model by $\sqrt{4\pi}$ to account for the bug, although a new improved version of the emulator will soon be trained that does not include the bug. Errata will be published for all of these papers.

It is also important to note that \rtdist is a very difficult model to work with because the parameter space is very complex, with strong correlations between several parameters. These correlations make it very difficult to find a reliable best fitting parameter combination, and give MCMC runs a very long autocorrelation length. The \reltransdcp parameters are less strongly correlated, and thus the model tends to be better behaved.} \\

\subsection{Future work}

In future we plan to include an extended corona into the \reltrans models in place of the lamppost. The first step of this process was the double lamppost model \rev{\citep{Lucchini2023}}, which we can eventually extend to $N$ lampposts followed by $N$ surface area elements of an extended corona. Each surface area element would not only have different light crossing delays due to different positions, but also the emission across all patches is expected to be non-uniform and fluctuating with time. These factors will introduce additional time-lags that are not currently considered within \reltrans models. The resulting model will be highly computationally intensive, necessitating the use of a neural network emulator \citep{Ricketts:2024}. The coronal parameters of such a model can then be constrained from reverberation mapping, folding in the known mass and distance.

Such an inference on coronal geometry can then be used on other objects without good existing mass and distance measurements. For BHXBs, this will be useful for probing the mass function in a way that circumvents the bias of dynamical measurements towards objects in low-extinction regions of the Galaxy, which may systematically contain less massive black holes \citep{Jonker:2021}. Since X-ray reverberation mapping does not suffer from the same bias, it could in principle be used to assess to what extent selection effects can account for gravitational wave sources appearing to host heavier black holes than BHXB systems \citep{Abbott:2023}. 

For AGN, X-ray reverberation mapping can in principle be used to measure the Hubble constant, $H_0$. \cite{Ingram:2022} calculate that a statistical uncertainty of $\sigma_{H_0} \sim 6~{\rm km} {\rm s}^{-1} {\rm Mpc}^{-1}$ can be achieved from a joint analysis of $\sim 25$ AGNs. This uncertainty is comparable to the discrepancy between the $H_0$ estimates derived the cosmic microwave background radiation and the traditional distance ladder. At the time of writing, a pilot study using \rtdist to model the X-ray reflection spectrum of the AGN Ark 564 is ongoing (Mitchell et al, in prep).
The above goals are of course highly ambitious. At the very least, we have shown that considering the mass and distance can provide a valuable consistency check on any model, and that accounting for prior knowledge of these two parameters can significantly aid inferences on the source geometry. 

This study has focused on RXTE data. Although the dataset is of very high quality, future work will benefit from utilizing more recent observations that cover a wider energy band pass. For example, simultaneous NICER and NuSTAR observations will enable us to cover the entire $\sim 0.5-75$ keV band pass. The higher energies covered by NuSTAR will provide better constraints on the Compton hump, whereas the soft X-ray coverage of NICER will increase our sensitivity to disk density -- which is key to the distance inference. A large caveat is that the soft X-rays are where the models are the most uncertain, whereas they are also where the count rate, and therefore signal to noise, is highest. Previous reverberation fits to NICER data (e.g. \citealt{Wang:2021}) have therefore likely been driven by soft X-rays rather than by the iron line and Compton hump. 

One problem is that soft X-rays are most sensitive to assumptions that influence the ionization state of the disc atmosphere. This includes the aforementioned seed photon temperature. Further to this, current reflection models assume that the disc atmosphere is only irradiated from above by the corona, and not from below by the thermal emission from the disc mid-plane. This assumption may hold for the hard state, but likely breaks down in the intermediate state. Moreover, the \rev{current} \texttt{xillver} tables are limited to disc densities $n_e \leq 10^{20}~{\rm cm}^{-3}$ \citep{Mastroserio:2021}. Although the best fitting density parameter is below this limit for all our models (see Table \ref{tab:params}), the parameter we fit for is the \textit{minimum} density in the disc as a function of radius. There will therefore be radii in our best fitting models with the density saturated at $10^{20}~{\rm cm}^{-3}$ instead of following the Zone A \cite{SS:1973} density profile. \texttt{reflionx} grids can extend to higher density \citep[e.g.][]{Tomsick:2018}, \revnew{though we note that this is because \texttt{reflionx} uses a less complete treatment of atomic physical processes than \texttt{xillver} \cite{Garcia:2010}}. Much of the relevant physics has recently been added to the \texttt{xillver} source code \citep{Ding:2024}, and new grids extending to higher densities will be available in the future. This will be hugely beneficial to reflection and reverberation analyses of X-ray binaries.

Another approximation important for soft X-rays is the implicit assumption that reflection responds instantaneously to changes in the illuminating flux. This is a very good assumption for the iron line and Compton hump \citep{Garcia:2013b}, whereas photons that emerge in the soft X-rays may do so after many interactions, which may have taken some significant thermalization time \citep{Salvesen:2022}. These caveats may help to explain why the soft lags often attributed to reverberation are observed to increase dramatically during the hard to soft transition \citep{Wang:2022} whereas the polarisation properties, and thus presumably the coronal geometry, remain remarkably constant \citep{Ingram:2024}.

\section{Conclusions}
\label{sec:conclusions}

We have fit X-ray reverberation models from the \reltrans package to RXTE data of Cyg X-1 in the hard state, yielding estimates of distance and black hole mass. We find evidence for disc variability that leads the continuum variability; which we interpret as the effect of inwardly propagating fluctuations in the disc mass accretion rate. \rev{For our best fitting model (Model B), we find $M=15\pm 4~M_\odot$ and $D=3.4_{-1.2}^{+1.6}$ kpc (with 90 per cent uncertainties), in agreement with recent dynamical and parallax measurements \citep{Ramachandran:2025}. Another model (Model C), employing different assumptions about the angular emissivity of the corona, returned a significantly worse fit ($\Delta \chi^2 = 48$ with two extra free parameters). Since Model C is essentially a self-consistent treatment of a lamppost corona, we essentially disfavour a geometry whereby the corona is very compact and located on the black hole spin axis, and is instead extended to some degree. In future, we plan to develop extended versions of \reltrans featuring an extended corona.}


\section*{Acknowledgments}

PO acknowledges support from STFC, AI acknowledges support from the Royal Society. PO would like to thank Javier Garc\'ia, Erin Kara for the support provided during the research and writing process and Chris Done and Jaichen Jiang for their useful comments and insights.   

\section*{Data Availability}
The observational data used in this research are public and available for download from the HEASARC. The \reltrans~\rev{source code} is publicly available \rev{at \url{https://github.com/reltrans/reltrans}. Documentation can be found at \url{https://reltransdocs.readthedocs.io/en/latest/}.}

\bibliographystyle{mnras}
\bibliography{main} 

\appendix

\section{Poisson noise in the presence of dead time}
\label{sec:pois}

Poisson noise \revnew{only contributes to the cross spectrum}
if the subject band is within the reference band (i.e. if $n_{\rm min} \leq n \leq n_{\rm max}$). We calculate the Poisson noise as
\begin{equation}
    P_{\rm noise}(E_n,\nu)=\begin{cases}
			2 \mu(E_n) w(\nu) & \text{if $n_{\rm min} \leq n \leq n_{\rm max}$,}\\
            0 & \text{otherwise.}
		 \end{cases}
\end{equation}
Here, $\mu(E_n)$ is the mean count rate in the $n^{\rm th}$ energy channel detected by all five PCUs. The final term departs from $w(\nu)=1$ if there is a non-zero deadtime $\tau_d$ after the detection of each photon during which it is not possible to detect another photon. We use the formula \citep{Vikhlinin:1994,Zhang:1995}
\begin{equation}
w(\nu) = 1 - 2 \frac{ [1-\cos(\omega \tau_d)] + (\omega/\mu_{\rm in}) \sin(\omega \tau_d) } { [1-\cos(\omega \tau_d)]^2 + [ \sin(\omega \tau_d) + \omega/\mu_{\rm in} ]^2 },
\end{equation}
where $\omega = 2\pi \nu$ and $\mu_{\rm in}$ is the mean incident count rate per PCU summed over all energy channels, which relates to the detected count rate per PCU $\mu_{\rm det}$ as
\begin{equation}
    \mu_{\rm in} = \frac{ \mu_{\rm det} }{ 1 + \tau_d \mu_{\rm det} }.
\end{equation}
We set $\mathcal{N}(E_n,\nu_j)$ as the Poisson noise averaged over the frequency range centred on $\nu_j$. We use $\tau_d = 10~\mu{\rm s}$ for the PCA deadtime \citep{Nowak:1999}. Since the deadtime is small, the Poisson noise is constant to a good approximation, with
\begin{equation}
w(\nu) \approx 1 - \frac{2 \tau_d \mu_{\rm in}}{1 + 2 \tau_d \mu_{\rm in}}.
\end{equation}

\rev{\section{Systematic errors}
\label{sec:sys}

If the specific photon flux from the source that is crossing the detector is $dN/dE$ (units of photons per unit area per unit time per unit energy), the count rate detected in the $I^{\rm th}$ energy channel is
\begin{equation}
    S(I) = \int_0^\infty A_I(E) ~\frac{dN}{dE}~dE,
    \label{eqn:SI}
\end{equation}
where $A_I(E)$, known as the instrument response, is the effective area of the $I^{\rm th}$ energy channel as a function of true photon energy $E$. The total effective area $A(E)$ is the sum of $A_I(E)$ over all channels. Our model for the instrument response, $\tilde{A}_I(E)$, is stored as a response matrix on a quantized photon energy grid, either as one file with extension \texttt{.rsp} (as is the case with RXTE), or as two files with extensions \texttt{.arf} and \texttt{.rmf} that contain $\tilde{A}(E)$ and $\tilde{A}_I(E)/\tilde{A}(E)$ respectively.

We typically have some systematic uncertainty in our estimate of the instrument response such that the true response relates to our model for it as $A_I(E) = \tilde{A}(E) \pm dA_I(E)$. It is common to assume this uncertainty to be some constant fraction $f$ of the true response, $dA_I(E) = f A_I(E)$. Subbing this into Equation (\ref{eqn:SI}), we see that the systematic uncertainty that this introduces to $S(I)$ is
\begin{equation}
    dS(I) = \int_0^\infty dA_I(E) ~\frac{dN}{dE}~dE = f S(I).
    \label{eqn:dSI}
\end{equation}
This is why it is so often appropriate to employ a constant fractional error in spectral fitting to account for systematic uncertainty in our model of the instrument response.

It is very easy to extend this formalism to the cross spectrum, due to the linearity of the Fourier transform. Let us define our model for the cross spectrum of the source as $G(E,\nu) = R^*(\nu) dN(E,\nu)/dE$, where $R(\nu)$ is the Fourier transform of the time-dependent count rate in the reference band, and $dN(E,\nu)/dE$ is the Fourier transform of the time-dependent specific photon flux as a function of (true) photon energy. The `observed' cross spectrum, $G(I,\nu) = S(I,\nu) R^*(\nu)$, where $S(I,\nu)$ is the Fourier transform of the time-dependent count rate in the $I^{\rm th}$ energy channel, can be written as \citep{Mastroserio:2018}
\begin{equation}
    G(I,\nu) = \int_0^\infty A_I(E) ~G(E,\nu)~dE.
    \label{eqn:GI}
\end{equation}
Therefore a systematic uncertainty in instrument response $dA_I(E)=f A_I(E)$ leads to a systematic uncertainty in the cross spectrum of
\begin{equation}
    dG(I,\nu) = \int_0^\infty dA_I(E) ~G(E,\nu)~dE = f G(I,\nu).
    \label{eqn:dGI}
\end{equation}
For our analysis, we use $f=4\times 10^{-3}$. Note that Equation (\ref{eqn:dGI}) is simply a generalisation of Equation (\ref{eqn:dSI}), since $S(I) \propto G(I,\nu=0)$.}

\section{Empirical parameters}
\label{sec:cross_spectrum}

\begin{figure*}
     \centering
     \includegraphics[width=\linewidth]{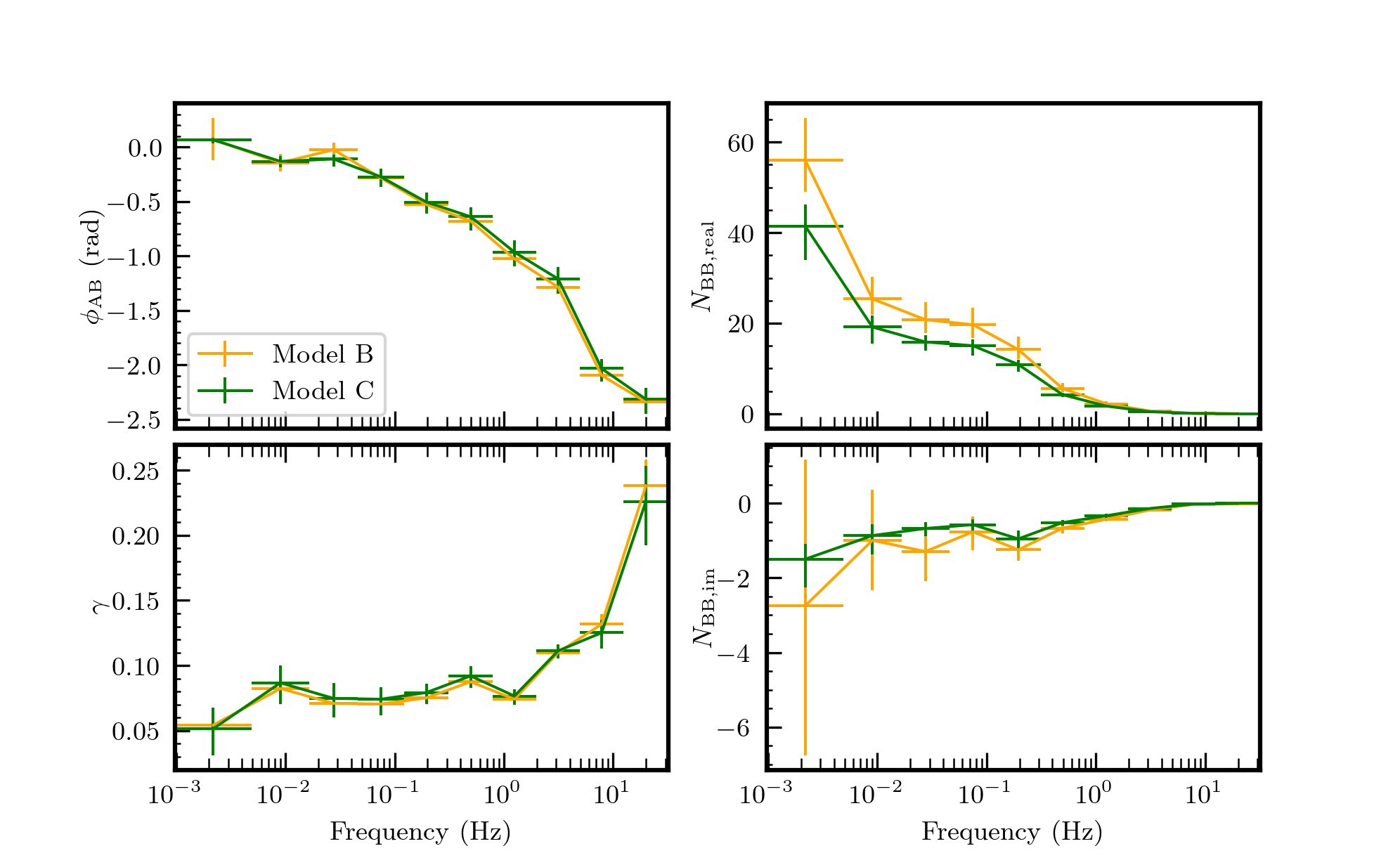}
     \vspace{-20pt}
     \caption[empirical]{Best fit empirical parameters $\phi_{AB}$ (\textit{top left}) and $\gamma$ (\textit{bottom left}); and \texttt{bbody} normalizations for the real (\textit{top right}) and imaginary parts (\textit{bottom right}) of the cross spectrum in each frequency band.}
     \label{fig:empirical}
\end{figure*}

Here we present the empirical modelling parameters, i.e. the normalization of the blackbody component for real and imaginary parts of the cross spectrum for each frequency range, and the continuum lag parameters $\phi_{\rm{AB}}$ and $\gamma$. In the \reltrans models, continuum lags are generated by spectral pivoting, where $\phi_{\rm{AB}}$ and $\gamma$ represent the phase difference and amplitude ratio between the spectral index and normalization variability respectively. For the cross spectrum in each frequency band, the spectral pivoting parameters are tied between real and imaginary parts during fitting. Fig. \ref{fig:empirical} (left) shows the best fitting continuum lag parameters for Models B and C.

These parameters have undergone redefinition since M19 \citep{Mastroserio:2021}. We use the \cite{Mastroserio:2021} redefinition of $\phi_{\rm{AB}} = \phi_{\rm{B}} +\pi -\phi_{\rm{A}}$, where $\phi_{\rm{A}}$, $\phi_{\rm{B}}$ were the previously implemented spectral pivoting parameters (in addition to $\gamma$) used in M19. A back-of-the-envelope calculation obtains M19 $\phi_{\rm{AB}}$ values that range from $\phi_{\rm{AB}} \sim 6.1$ radians, at low frequencies, to $\phi_{\rm{AB}}\sim3.9$, rad at high frequencies. For Models B and C here, we find a similar range of values (i.e. $\sim 2$) across the full frequency range. The difference in the measured $\phi_{\rm{AB}}$ values compared with M19 relates to the different of reference band used. Unlike M19, our measured $\gamma$ increases with frequency. It is unclear whether the \cite{Mastroserio:2021} redefinition of $\gamma$ parameter would produce such a trend in our observed $\gamma$, but considering we are using a new cross-spectral error formula it is not unexpected that these parameters take different values. The inclusion of disc variability will also influence the best-fitting continuum lag parameters, since the disc now dominates the low-frequency lags that were previously modelled only by spectral pivoting in the M19 fit.
Disc variability is represented by the \bbody normalization $N_{\rm{BB}}$, which is treated as free parameter for the real and imaginary part of the cross spectrum in each frequency band (i.e. $N_{\rm {BB}, \rm{real}}$ is independent of  $N_{\rm {BB}, \rm{im}}$ for each frequency range). The best fitting values for Models B and C are plotted in Fig. \ref{fig:empirical}. These values are used to calculate the lag and variability amplitude presented in Fig. \ref{fig:thermal_lags}.

\revnew{\section{Variability amplitude of the variable disc region}
\label{sec:discrms}

We approximate the time-dependent bolometric flux from the disc crossing our detector as
\begin{equation}
    F(t) = F_c + a(t) F_v,
\end{equation}
where the constant component originates from the radial range $r=r_v \rightarrow \infty$, the variable component from the range $r=r_{\rm in} \rightarrow r_v$, and the dimensionless function $a(t)$ has a mean of unity. We fit a \texttt{bbody} component to the cross spectrum to approximate the variable disc region. The flux from the variable region $F_v$ relates to the Fourier frequency-dependent \texttt{bbody} normalisation $N_{\rm BB}(\nu_j)= \sqrt{N_{\rm BB,real}^2(\nu_j) + N_{\rm BB,im}^2(\nu_j)}$ as
\begin{equation}
    |A(\nu_j)| F_v = \frac{ 55.19~N_{\rm BB}(\nu_j)~{\rm keV}~{\rm cm}^{-2}~{\rm s}^{-1} }{ \sqrt{P_r(\nu_j)}},
    \label{eqn:FvAnu}
\end{equation}
where $P_r(\nu_j)$ is the power spectrum of the reference band count rate in absolute rms normalisation and $A(\nu_j)$ is the Fourier transform of $a(t)$. The total fractional variance of the variable region is
\begin{equation}
    {\rm rms}^2 = \sum_{j=1}^{N_f} |A(\nu_j)|^2 \Delta\nu_j,
    \label{eqn:rms1}
\end{equation}
where $\Delta \nu_j$ is the width of the $j^{\rm th}$ frequency range used in our fit, and $N_f=10$ in our case.

We can therefore constrain the variance of the variable region, $F_v^2 \times {\rm rms}^2$, from the \texttt{bbody} parameters alone, but to constrain the \textit{fractional} variance we need an independent constraint of the flux of the variable region $F_v$. For this, we can use the \texttt{diskbb} fit to the time-averaged spectrum. The total disc flux relates to the \texttt{diskbb} normalisation $N_d$ as
\begin{equation}
    F = 0.018~N_d \left( \frac{kT_{\rm in}}{{\rm keV}} \right)^4 \int_0^1 y^{1/3}~dy~{\rm keV}~{\rm cm}^{-2}~{\rm s}^{-1},
\end{equation}
where $y=T/T_{\rm in}$. Assuming that the variable region covers the disc radii with temperature $T = T_v \rightarrow T_{\rm in}$, the flux of the variable region is
\begin{equation}
    F_v = \frac{3}{4} 0.018~N_d \left( \frac{kT_{\rm in}}{{\rm keV}} \right)^4 ( 1 - y_v^{4/3} )~{\rm keV}~{\rm cm}^{-2}~{\rm s}^{-1},
    \label{eqn:Fv}
\end{equation}
where $y_v = T_v/T_{\rm in}$. Combining Equations (\ref{eqn:FvAnu}), (\ref{eqn:rms1}) and (\ref{eqn:Fv}) gives
\begin{equation}
    {\rm rms} = \frac{ 4 / 3 }{ K_d (kT_{\rm in}/{\rm keV})^4 ( 1 - y_v^{4/3} ) } \sqrt{\frac{\sum_j N_{\rm BB}^2(\nu_j) \Delta\nu_j}{P_r(\nu_j)}}.
\end{equation}
In our best fit, we find $K_d=700$, $kT_{\rm in}=0.71$ keV, and $kT_v=0.64/1.35~{\rm keV} = 0.47$ keV. Subbing in the measured reference band power spectrum and the best-fitting Model B \texttt{bbody} normalisations (Fig \ref{fig:thermal_lags} bottom, yellow) gives rms $\approx 45\%$. The total disk flux for these parameters is $F \approx 2.4~{\rm keV}~{\rm cm}^{-2}~{\rm s}^{-1}$, whereas the flux of the variable region is $\approx 1.0~{\rm keV}~{\rm cm}^{-2}~{\rm s}^{-1}$.}

\label{lastpage}
\end{document}